\documentclass[10pt, aps,pre, twocolumn,superscriptaddress]{revtex4-2}
\usepackage{amsmath,amssymb,amsfonts,bbm,graphicx,times,psfrag}
\usepackage[pdftex]{color}
\usepackage{graphicx}
\usepackage[colorlinks, linkcolor=blue, citecolor=blue, urlcolor=blue, breaklinks=true]{hyperref}
\usepackage{braket}
\usepackage{soul}
\usepackage{autobreak}
\usepackage{microtype}

\usepackage{amsthm}
\usepackage{csquotes}
\usepackage{amsthm}\newcommand{\be} {\begin{equation}}
\newcommand{\ee} {\end{equation}}

\usepackage{color}
\usepackage{mathrsfs}
\usepackage{amssymb}

\usepackage[skip=2pt]{caption}
\usepackage[utf8]{inputenc}
\DeclareUnicodeCharacter{2283}{\supset}
\usepackage{ragged2e}
\begin{document}
\title{Controllable diatomic molecular quantum thermodynamic machines}
\author{C.O. Edet}
\email{collinsokonedet@gmail.com}
\affiliation{Institute of Engineering Mathematics, Universiti Malaysia Perlis, 02600 Arau, Perlis, Malaysia}
\affiliation{Department of Physics, University of Cross River State, Calabar, Nigeria}
\author{E. P. Inyang}
\affiliation{Department of Physics, National Open University of Nigeria, Jabi, Abuja}

%\author{P. O. Amadi}
%\affiliation{Institute of Engineering Mathematics, Universiti Malaysia Perlis, 02600 Arau, Perlis, Malaysia}
\author{O. Abah}
%\email{obinna.abah@newcastle.ac.uk}
\affiliation{School of Mathematics, Statistics and Physics, Newcastle University, Newcastle upon Tyne NE1 7RU, United Kingdom}
\author{N. Ali}
\affiliation{Faculty of Electronic Engineering \& Technology, Universiti Malaysia Perlis, 02600, Arau, Perlis, Malaysia  }
\begin{abstract}
We present quantum heat machines using a diatomic molecule modelled by a $q$-deformed potential as a working medium. We analyze the effect of the deformation parameter and other potential parameters on the work output and efficiency of the quantum Otto and quantum Carnot heat
cycles. Furthermore, we derive the analytical expressions of
work and efficiency as a function of these parameters. 
Interestingly, our system operates as a quantum heat engine across the range of parameters considered.
In addition, the efficiency of the quantum Otto heat engine is seen to be tunable by the deformation parameter.  Our findings provide useful insight for understanding the impact of anharmonicity on the design of quantum thermal machines. 
\end{abstract}
\maketitle

\section{Introduction}
Thermal machines have been at the heart of thermodynamics since the early days of the field \cite{cengel2011thermodynamics}. They are often harnessed to provide mechanical energy from thermal energy. Macroscopic
motors, refrigerators, and heat pumps are notable examples of classical machines whose dynamics follow the laws
of classical physics \cite{cengel2011thermodynamics}.  
The advances in nanofabrication have led to the development of smaller devices. On this scale, the quantum effects dominate, and the laws must be considered. Since the 1950s, there have been intense theoretical endeavours in comprehending the interplay of thermodynamics and quantum physics following the seminal paper Scovil and Schulz-DuBois on maser heat engines \cite{scovil1959three}. 

Quantum heat engines produce work using quantum matter as a working substance and exhibit exotic properties. They have been shown to perform better compared to their classical counterpart \cite{myers2022quantum}. In recent times, several quantum mechanical models have been proposed as a working medium, for example, the harmonic oscillator potential \cite{quan2007quantum}, infinite square well potential \cite{quan2007quantum}, spin systems \cite{ryan2008spin,friedenberger2017quantum,das2019measurement, huang2014quantum}, stanene \cite{fadaie2018topological}, strained-graphene \cite{mani2017strained}, Dirac particles \cite{munoz2012quantum, pena2016optimization}, two-level system \cite{kieu2006quantum}, multi-level system \cite{quan2005quantummulti}, a continuum working medium \cite{li2007quantum}, photon gas \cite{hardal2015superradiant}, diatomic molecules \cite{oladimeji2024performance}, Posch-Teller potential \cite{oladimeji2019efficiency,oladimeji2021poschl}, among others. In addition, the past decade has witnessed several experimental implementations of quantum thermal machines \cite{rossnagel2016single, myers2022quantum, blickle2012realization, von2019spin, peterson2019experimental, klatzow2019experimental, passos2019optical}.

Moreover, quantum systems that exhibit anharmonicity, such as molecular systems \cite{morse1929diatomic, schmidt2024molecular}, superconducting qubits \cite{krantz2019quantum}, are of great interest in the development of quantum technologies. Anharmonic effects have been shown to be useful in several settings \cite{home2011normal}, but most quantum systems are often treated with approximate harmonic theoretical descriptions, for instance in the case of a single molecule \cite{galperin2007molecular,lu2015effects, lu2011laserlike, simine2012vibrational, simine2013path,
 arrachea2014vibrational,agarwalla2016reconciling, erpenbeck2015effect}. The harmonic approximation is valid when atomic displacements are limited \cite{friedman2017effects}, and recent experiments are beginning to probe the limits of this model in a plethora of processes and contexts \cite{xomalis2021detecting,chen2021continuous,lombardi2018pulsed}. However, most quantum thermal machines assume that their working mediums are often harmonic in nature \cite{Kosloff2017,rossnagel2016single,myers2022quantum}, but recent experiments show the possibility of trapping molecules in anharmonic traps \cite{home2011normal}. Therefore, it becomes crucial to develop quantum thermal machines with genuine anharmonicity to see how they impact the performance of thermal machines. In this regard, a Carnot-like engine with a diatomic molecule has recently been proposed by \citet{oladimeji2024performance}. Here, we extend this to carry out a full quantum thermodynamic description of a quantum thermal machine whose working medium is a tunable Morse potential, which gives us the ``degree of freedom" to tune the behaviour of the model. The Morse potential is a well-known diatomic molecular potential that captures the effect of anharmonicity and real bond-breaking \cite{morse1929diatomic}.   
We investigate the anharmonic effects on the performance of a quantum thermal machine in an analytically feasible model using a generalized  ($q$-deformed) Morse potential as the working substance. The choice of this model is inspired by the fact the presence of the $q$-parameter gives us tunable control in our setup. %The structure of this system is schematically presented in Fig (\ref{figmodel}) (detail of the system is presented in the following section). 
This $q$-deformed physical system is typically described by $q$-deformed potentials which is reduced to their normal equivalent when $q\rightarrow 1$.  The $q$-deformed hyperbolic potentials first appeared in the seminal works of \citet{arai1991exactly, arai2001exact}. This class of model potentials has since found applications in several areas of research, notably the study of the phonon spectrum in ${}^4\text{He}$ \cite{r1996quantum}, multi-atomic molecules \cite{bonatsos1997coupled}, oscillatory-rotational spectra of diatomic \cite{johal1998two},  the description of the electronic conductance in disordered metals and doped semiconductors \cite{alavi2004exact} amongst many other\cite{cooper1995q, ikhdair2009rotation, dobrogowska2013q}. A recent study showed that statistical complexity on the information-theoretic measure of diatomic molecules is strongly influenced by the potential deformation parameter $q$ \cite{nutku2022complexity}. Previous studies of quantum heat engines with Morse potentials have focused on the undeformed case \cite{oladimeji2024performance}. Separately, q-deformed oscillators have been explored as effective models for anharmonic quantum systems. In contrast, the present work combines these two directions by introducing a q-deformed Morse potential as the working medium of a quantum heat engine. The deformation parameter $q$ provides an additional control knob that modifies the vibrational spectrum, enabling tunable a quantum heat engine.

%Therefore, it is the goal of this study to extend the application of such a physical system to quantum thermodynamics.

In this paper, we study the performance of a quantum thermal machine using a diatomic molecule as a working substance. We show that the presence of the deformation parameter and other potential parameters can be used to regulate the behaviour of the system such that different operational phases are revealed in different parameter regimes.
This article is organized as follows. We introduce the diatomic molecule model in Sec. \ref{model}. In Sec. \ref{theory}, we present the theory and analysis of the quantum thermodynamic Carnot and the quantum Otto cycle of the controllable molecule model. In Sec. \ref{conclusion}, we present concluding remarks and discuss experimental feasibility
%%%%%%%%%%%%%%%%%%%%%%%%%%%%%%%%%%%%%%%

\section{The diatomic molecule}\label{model}
We begin by describing the model and previous results from the literature that are central to our analysis.
\begin{figure}[!t]
\centering
\includegraphics[width=0.9 \columnwidth]{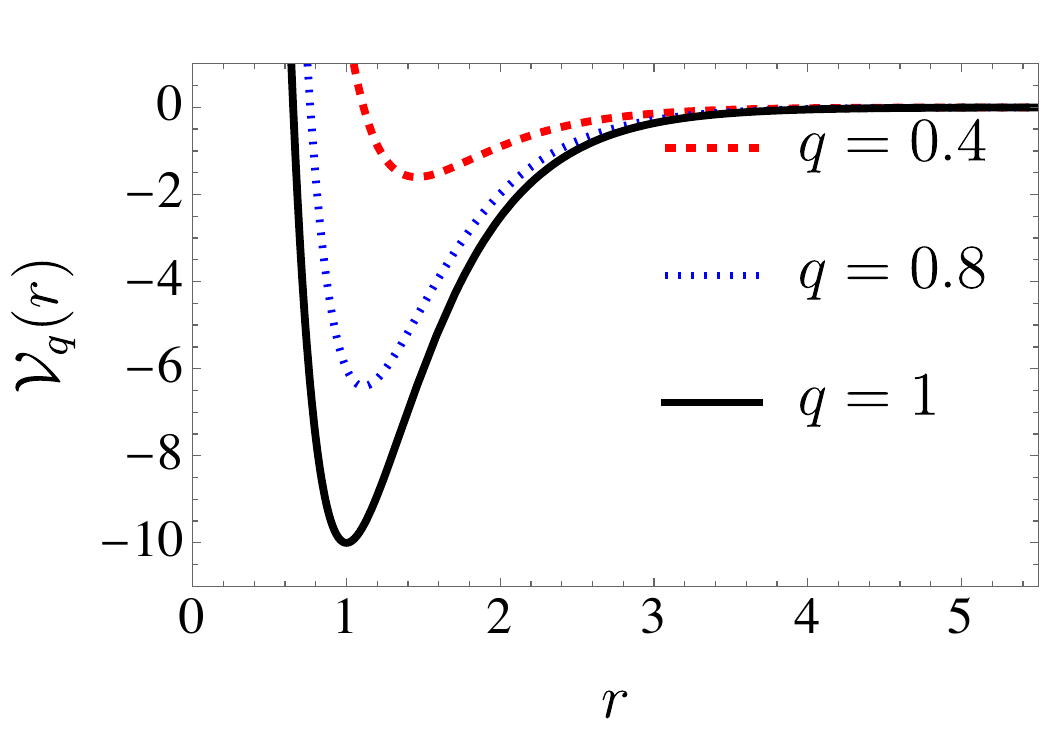} 
\caption{\justifying Plot of the  q-deformed Morse potential for $q\!\in\!\{0.4, ~0.5,~ 1\}$, where $q\rightarrow1$ corresponds to the standard Morse potential. Parameters used: $\mathcal{D}_{\text{e}}\!=\!10 $, and $\alpha\!=\!2 $. }       
\label{figmodel}
\end{figure}
The Hamiltonian describing the diatomic molecule system reads \cite{aktacs2004supersymmetric, ikhdair2009rotation, nutku2022complexity, hassanabadi2018superstatistics, boumali2018statistical},
%\begin{equation}\label{hamiltonian}
%H= \frac{p_{x}^2}{2 \mu r_e^2} + \mathcal{V}_q(x)
%\end{equation}
\begin{eqnarray}\label{hamiltonian}
H&=& \frac{p_{x}^2}{2 \mu r_e^2} + \mathcal{V}_q(x), \nonumber\\ &=& \frac{p_{x}^2}{2 \mu r_e^2} + \mathcal{D}_e (e^{-2 \xi x}-2  q e^{-\xi x}),
\end{eqnarray}
where the first term is the kinetic energy operator with $p_{x}\!=\!-i\hbar d/dx$ as the momentum operator, and $d/dx$  is the differential operator. The second term is the deformed Morse potential model $\mathcal{V}_q(x)$, where $x\!=\!(r-r_e)/r_e$, and $\xi =\alpha\,r_e. $ The first term of the potential corresponds to the repulsive part, and the second terms correspond to the attractive part of the potential. Here, $\mathcal{D}_e > 0$ is the potential strength or dissociation energy, it quantifies the depth of the potential well at the equilibrium distance, $r_e$ is the equilibrium bond length, $\alpha$ determines the width of the potential well, and $q$ is a dimensionless parameter for controlling the shape of the potential (see, Fig. (\ref{figmodel}). The  deformation parameter $q$, it is a real number $q \in  [0, 1]$. We recover the standard Morse potential \cite{morse1929diatomic} when $q=1$.  This Morse potential has been noted to be a suitable potential model for modeling the inter-atomic interaction in a diatomic molecule. This is because it accounts for the impact of bond breaking, such as the existence of unbound states. More so, it also accounts for the anharmonicity of real bonds. On expansion of $\mathcal{V}_q(x)$ from Eq. \eqref{hamiltonian} in a power series of $x$, we get 
\begin{equation}
 \mathcal{V}_q(x) \approx   D_e(1-2 q)+2 \xi  D_e (q-1)x+ \xi ^2 D_e\left(2-q\right) x^2+O\left(x^3\right)
\end{equation}
%It is interesting to note that in contrast to the standard Morse potential, $q$-deformed Morse potential cannot be approximated by the quantum harmonic oscillator except in the limit of $q \rightarrow 1$, $ \mathcal{V}_q(x)\!\approx\!-D_e+ \xi ^2 D_e x^2+O\left(x^3\right)$ in the neighbourhood of its minimum.  
For $q\neq 1$, the $q$-deformed Morse potential attains its minimum at $x=x_0=\ln(q)/\xi$ rather than at $x=0$. As a result, a Taylor expansion about $x=0$ contains a nonzero linear term and does not correspond to a harmonic approximation, whereas expanding about the true minimum $x_0$ yields a quadratic leading-order term and thus a well-defined harmonic approximation. Only in the limit $q\to 1$ does $x_0\to 0$, so that the expansion about $x=0$ coincides with the harmonic approximation. Hence, we can see that for small values of $x$ the Morse potential, Eq. \eqref{hamiltonian}, can be approximated by a harmonic oscillator with spring
constant $k = 2 \xi ^2 D_e$, and $V_q(x)\!\approx\!-D_e+ k/2 x^2$, this was illustrated by \citet{leonard2015quantum}. 
%It is observed that the Morse potential significantly deviates from the harmonic approximation as the distance from $x\!=\!0$ increases. This allows us to model bond weakening and dissociation with diverging nuclear separation
Although the Morse potential admits continuum states corresponding to dissociation, in this work we focus on its bound eigenstates, which already encode bond weakening through anharmonicity and the existence of a highest bound level. Moreover, the presence of the deformation reveals other features such as symmetry distortion \cite{nutku2022complexity}. This is particularly evident in the regime where the $q$ parameter is very small. The consequence of these properties introduces interesting features into the quantum heat engine performance, as we will discuss shortly in Sec. \ref{theory}. 

%($H \psi_n = \mathcal{E}_n \psi_n$) 
The solutions of the Schrodinger equation for the Hamiltonian, Eq. \eqref{hamiltonian}, yield the energy eigenvalue (see Appendix \ref{Solutionsappend} for details): 
\begin{align}\label{eigenvalue}
    \mathcal{E}_n=-\xi ^2 p \left(-n+\lambda  q-\frac{1}{2}\right)^2,
\end{align}
%$\mathcal{E}_n=-\xi ^2 p \left(-n+\lambda  q-\frac{1}{2}\right)^2$, 
where $n\!=\!0,1,2,3,..., n_{\text{max}}$, $\lambda \!=\!\sqrt{\frac{D_e}{\xi ^2 p}}$, and $p\!=\!\frac{\hbar ^2}{2 \mu  r_e^2}$.  The energy level associated with the $q$-deformed Morse oscillator is bounded from above by $q$  with the maximum quantum number given as $n_{\text{max}}\leq (\lambda q-1/2) $. 
The equilibrium thermodynamic behaviour of the $q$-deformed Morse potential can be determined from the canonical partition function,
\begin{equation}\label{parti4f}
    \mathcal{Z}= \text{Tr}[\text{exp} (-\beta H)]=\sum\limits_{n=0}^{n= \infty }  \text{exp}(-\beta \mathcal{E}_n),
\end{equation}
where $\beta\!=\!1/(k_\mathrm{B}T)$ is the inverse temperature $T$ with the Boltzmann constant $k_\mathrm{B}$. It is worth noting that  although the upper limit of the sums is formally written as $\infty$, only bound vibrational levels up to $n_{\text{max}}$ contribute; terms beyond vanish identically.In the energy representation, it is straightforward to calculate the partition function using  the $q$-deformed Morse potential eigen energies \eqref{eigenvalue}, we obtain
\begin{equation}\label{partitionfunction}
 \mathcal{Z} =   \frac{\sqrt{\pi } \text{erfc}\left(\frac{1}{2} (1-2 \lambda  q) \sqrt{\beta  \xi ^2 (-p)}\right)}{2 \sqrt{\beta  \xi ^2 (-p)}}.
\end{equation}
%This expression will be useful for our analysis of the quantum heat engine. It is worth noting that the partition function \eqref{partitionfunction} obtained above can be used to determine the equilibrium thermodynamic behavior of the q-deformed Morse Oscillator, such as the internal energy, free energy, entropy, and heat capacity \cite{huang2009introduction, callen1991thermodynamics}. 
 %It is important to note that these studies were conducted within the framework of equilibrium thermodynamics. 
The thermal state density matrix can be determined from the eigenstate and the partition function as,
\begin{align}
    \rho = \sum\limits_n \mathcal{P}_n |n\rangle \langle n| = \sum\limits_n (\text{exp}(-\beta \mathcal{E}_n)/ \mathcal{Z} ) |n\rangle \langle n|,
\end{align}
where $\mathcal{P}_n$ is the occupation probability of the $n^{\text{th}}$ eigenstate and $\mathcal{E}_n$ is the $n^{\text{th}}$ eigen energy of the working medium, $q$-deformed Morse potential.

 Hence, the total internal energy reads $\mathcal{U}\!=\!\sum\limits_n \mathcal{P}_n \mathcal{E}_n$, and the infinitesimal change yields $d\mathcal{U}\!=\!\sum\limits_n (\mathcal{E}_n d\mathcal{P}_n+\mathcal{P}_n d\mathcal{E}_n)$. Thus, the first law of thermodynamics can be deduced as, 
\begin{equation}
    d\mathcal{U}\!=\!d \mathcal{Q}+d \mathcal{W},
\end{equation}
where $d\mathcal{Q}\!=\!\sum\limits_n \mathcal{E}_n d\mathcal{P}_n$ and $ d\mathcal{W}\!=\!\sum\limits_n  \mathcal{P}_n d \mathcal{E}_n$. 
%$d\mathcal{U}\!=\!d \mathcal{Q}+d \mathcal{W}$. 
The internal energy, $\mathcal{U}$, is the state function, whereas heat $ \mathcal{Q}$ and work $\mathcal{W}$ are path-dependent functions. 
We remark that recent studies have focused on analyzing the equilibrium thermodynamic properties of the $q$-deformed Morse potential within the framework of superstatistics \cite{beck2003superstatistics,hassanabadi2018superstatistics} and the Euler-Maclaurin method \cite{boumali2018statistical}.
%\text{Tr}[\rho H]=
%%%%%%%%%%%%%%%%%%%%%%%%%%%%%%%%%%%%%%%%%%%%%%%%%%%%%%%
%\section{RESULTS AND DISCUSSION} \label{theory}
\section{Diatomic molecular quantum heat engine}\label{theory}
An aggregate of different quantum thermodynamic processes and the number of strokes per cycle yields several kinds of quantum heat engines \cite{quan2005quantum, quan2009quantum}. In this section, we present the thermodynamics analysis of the performance of the diatomic molecule engine for two distinct thermodynamic cycles, namely the Carnot and Otto cycles. 
%%%%%%%%%%%%%%%%%%%%%%%%%%%%%%%%%%%%%%%%%%%%%%%%%%%%%%%
\subsection{Quantum Carnot Cycle}
\begin{figure}[!]
    \centering
    \includegraphics[width=0.8 \linewidth]{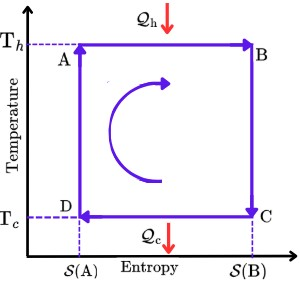}
    \caption{\justifying A pictorial representation of a Carnot thermal machine using a $q$-deformed Morse oscillator as a working substance. The engine cycle consists of two adiabatic strokes ($B \to C$ and $D \to A$) and two isothermal strokes, $A \to B$ at $T_{\text{h}}$ and $C \to D$ at $T_{\text{c}}$, with $T_{\text{h}} > T_{\text{c}}$.}
    \label{figqcarnot}
\end{figure}

Similar to the classical Carnot cycle, the cycle is reversible, and the energy eigenvalues must satisfy the reversibility condition for all states \cite{li2007quantum}. This cycle comprises four strokes, see Fig. \ref{figqcarnot}, namely; isothermal expansion, adiabatic expansion, isothermal compression, and adiabatic compression. 
The strokes $\text{B} \rightarrow \text{C}$ and $\text{D} \rightarrow \text{A}$ are adiabatic processes, while $\text{A} \rightarrow \text{B}$ and $\text{C} \rightarrow \text{D}$  are isothermal processes. During adiabatic strokes, $\mathcal{P}_n (\text{C})\!=\!\mathcal{P}_n (\text{B}) $, and $\mathcal{P}_n (\text{A})\!=\!\mathcal{P}_n (\text{D})$; that means $\mathcal{S} (\text{C})\!=\!\mathcal{S} (\text{B})$, and $\mathcal{S}(\text{A})\!=\!\mathcal{S}(\text{D})$.
%%%%%%%%%%%%%%%%%%%%%%%%%%%%%%%%
The heat exchanges during the cycle are given as; $\mathcal{Q}_{\text{h}}\!=\!T_{\text{h}}[\mathcal{S} (\text{B})-\mathcal{S}(\text{A})]>0$, and $\mathcal{Q}_{\text{c}}\!=\!T_{\text{c}}[\mathcal{S} (\text{D})-\mathcal{S} (\text{C})]<0$. The total work output reads
\begin{equation}\label{eq06}
\mathcal{W}^{\text{QC}} = Q_{\text{h}} + Q_{\text{c}} = (T_{\text{h}} - T_{\text{c}})\left[S(\text{B})-\mathcal{S}(\text{D}))\right],
\end{equation}
where $\mathcal{S}(i)$ is the entropy given as
\begin{equation}\label{entropyexp}
\mathcal{S}(i)= -k \sum \frac{\exp[-\beta_i \mathcal{E}_n(i)]}{\mathcal{Z}(i)}[-\beta_i \mathcal{E}_n(i)-\text{ln}\mathcal{Z}(i)],
\end{equation}
with  $i\in \{\text{A}, \text{B}, \text{C}, \text{D}\}$, the inverse temperature $\beta\!=\!1/k_\text{B}T$, and $k_\text{B}$ is the Boltzmann constant. To realize a quantum heat engine, $\mathcal{W}>0$ must be satisfied \cite{Prakash2022, pal2022josephson}. Employing the energy eigenvalue, $\mathcal{E}_n$, the entropy is evaluated using the approximation $\sum\limits_{n=1}^{\infty} \exp(-\beta \mathcal{E}_n) \approx \int_{0}^{\infty} \exp(-\beta \mathcal{E}_n)dn$ and reads 
\begin{equation}
\mathcal{S}_{h(c)} = k \log \left(\frac{\sqrt{\pi } \text{Erfc}\left(\gamma_1^{h(c)}\right)}{2 \sqrt{\beta_{h(c)}  \xi_{h(c)} ^2 (-p)}}\right)-\frac{1}{2} k \left( \Gamma_1 -1\right),
\end{equation}
where $\gamma_1^{h(c)}\!=\!\frac{1}{2} (1-2 \lambda_{h(c)}  q) \sqrt{\beta_{h(c)}  \xi_{h(c)} ^2 (-p)}, \gamma_2^{h(c)}= (2 \lambda_{h(c)}  q-1) \sqrt{\beta_{h(c)}  \xi_{h(c)} ^2 (-p)}$, $\gamma_3^{h(c)}\!=\!\frac{1}{4} \beta_{h(c)} \xi_{h(c)}^2 p (1-2 \lambda_{h(c)}q)^2$, and $\Gamma_1 = \left(\gamma_2^{h(c)} e^{\gamma_3^{h(c)}}\right)/\left(\sqrt{\pi } \text{Erfc}\left(\gamma_1^{h(c)}\right) \right)$.

%%%%%%%%%%%%%%%%%%%%%%%%%%%%%%%%%%%%%%%%%%%%
\begin{figure*}[!ht]
\centering
\includegraphics[width=1.9 \columnwidth]{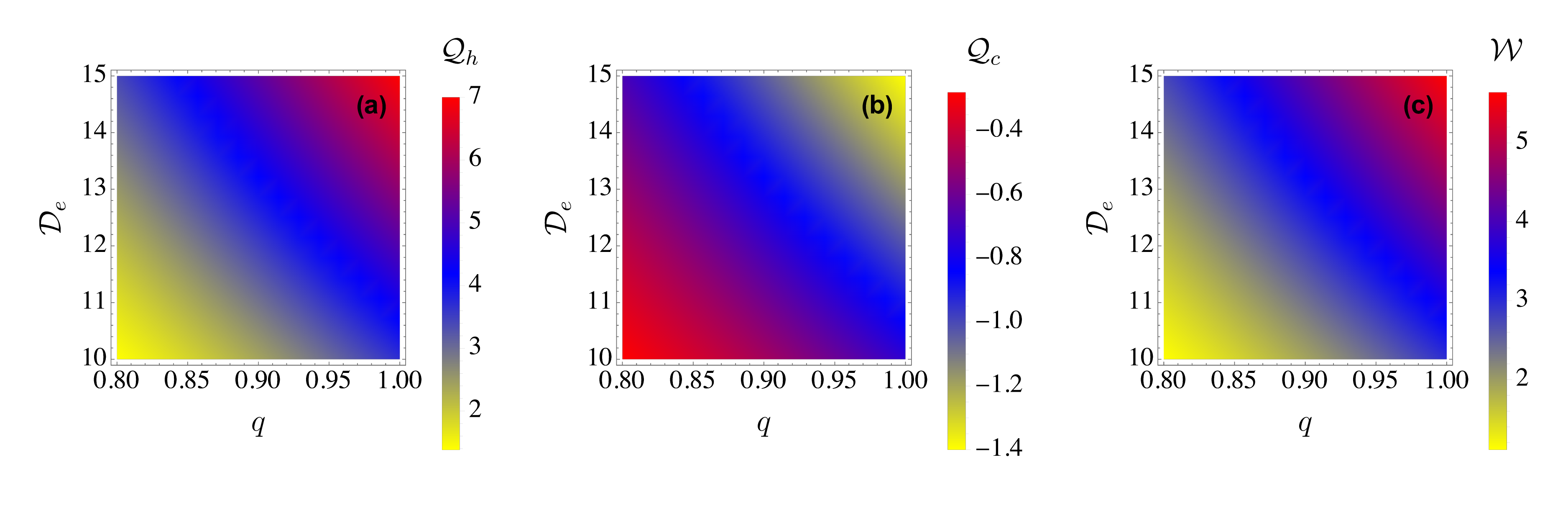} 
\caption{ \justifying Quantum Carnot engine: (\textbf{a}) The heat exchanges of the working substance with the hot reservoir,  $\mathcal{Q}_{\text{h}}$ vs $\mathcal{D}_{e}$ and $q$ (\textbf{b}) The heat exchanges of the working substance
with the cold reservoir,  $\mathcal{Q}_{\text{c}}$ vs $\mathcal{D}_{e}$ and $q$. (\textbf{c}) The Work output   $\mathcal{W}$ vs $\mathcal{D}_{e}$ and $q$. Parameters used: $\alpha_h = 2.236 $, $\alpha_c = 1$, $T_c\!=\!2 $ and $T_h = 10 $ }   
\label{fig1QCC}
\end{figure*}
%%%%%%%%%%%%%%%

The reversibility condition  is defined as \cite{li2007quantum}:
\begin{align}
\mathcal{E}_{\text{n}}(\text{C})- \mathcal{E}_{\text{m}}(\text{C}) = \frac{T_{\text{c}}}{T_{\text{h}}} [\mathcal{E}_{\text{n}}(\text{B})- \mathcal{E}_{\text{m}}(\text{B})],
\end{align}
where $\mathcal{E}_{n(m)}$ is the $q$-deformed Morse potential energy eigenvalue. The reversibility condition is satisfied independent of the states $n(m)$ and the variation in the width ($\alpha$) of the $q$-deformed Morse potential while other parameters such as the deformation parameter $q$, the dissociation energy $\mathcal{D}_e$, and the equilibrium bond length $r_e$ are kept constant during the cycle. Thus, if $\sqrt{\mathcal{D}_{e}^\text{C}/\alpha_\text{C}^2}\!=\!\sqrt{\mathcal{D}_{e}^\text{B}/\alpha_\text{B}^2}$, we obtain
\begin{equation}\label{reversibility}
\frac{\alpha_{\text{C}}^2}{\alpha _{\text{B}}^2}=\frac{T_c}{T_h},
\end{equation}
as the reversibility condition. Equation \eqref{reversibility} shows that the reversibility condition is satisfied by taking appropriate values of the width of the $q$-deformed Morse potential. Now, using the entropies within the reversibility condition, the work done for a quantum Carnot engine whose working medium is a $q$-deformed Morse oscillator can be expressed as; 
\begin{widetext}
\begin{equation} \label{eqWQC}
  \mathcal{W}^{\text{QC}} = \frac{k (T_\text{c}-T_\text{h}) \left(\text{Erfc}\left(\gamma_1^{c}\right) \left(2 \sqrt{\pi } \text{Erfc}\left(\gamma_1^{h}\right) \left(\log \left(\frac{\text{Erfc}\left( \gamma_1^{c}\right)}{\sqrt{\beta_c \xi_c^2 (-p)}}\right)-\log \left(\frac{\text{Erfc}\left(\gamma_1^{h}\right)}{\sqrt{\beta_h \xi_h^2 (-p)}}\right)\right)+\gamma^{h}_2 e^{\gamma_{3}^{h}}\right)-\gamma_2^ce^{\gamma_3^{c}} \text{Erfc}\left(\gamma_1^h\right)\right)}{2 \sqrt{\pi } \text{Erfc}\left(\gamma_1^c\right) \text{Erfc}\left(\gamma_1^{h}\right)}.
\end{equation}
\end{widetext}
The corresponding efficiency reads,
\begin{equation}
    \eta^{\text{QC}} = \frac{W^{\text{QC}}}{\mathcal{Q}_{\text{h}}} =\frac{(T_{\text{h}}-T_{\text{c}})(\mathcal{S}_{\text{h}}-\mathcal{S}_{\text{c})}}{T_{\text{h}}(\mathcal{S}_{\text{h}}-\mathcal{S}_{\text{c})}}= 1 - \frac{T_{\text{c}}}{T_{\text{h}}}.
\end{equation}

%%%%%%%%%%%%%%%%%%%%%%%%%%%%%%%%%%%%%%%%%%%%%%%%%%%%%%%%%%%% 
In Fig. (\ref{fig1QCC}), we present the numerical results of the heat exchanges and the total work extracted against the control/deformation parameter ($q/r_e$) and the dissociation energy ($D_e/r_e$). Fig. \ref{fig1QCC} (\textbf{a}) shows that the input heat is positive, that is $\mathcal{Q}_{\text{h}}>0$ while Fig. \ref{fig1QCC}(\textbf{b}) demonstrate that the output heat is always negative $\mathcal{Q}_{\text{c}}<0$ within the parameters range. In Fig. \ref{fig1QCC}(\textbf{c}), we observe an increase in work output as both $q/r_e$ and dissociation energy $D_e/r_e$ increase. Both parameters enhance the work output but have a stronger effect when they are at their maximum. {This demonstrates that the width of the well, controlled by $\alpha$, and the hot bath temperature both play crucial roles in enhancing the work output: the former governs the expansion and compression of the working substance, while the latter determines the thermal population of excited vibrational states during the isothermal process. Thus, considering that the total work output is positive, we conclude that the system operates in the quantum heat engine regime under the conditions studied.  The efficiency of the quantum Carnot $q$-deformed Morse oscillator heat engine is 0.8 at temperatures that satisfy the reversibility condition.  
%%%%%%%%%%%%%%%%%%%%%%%%%%%%%%%%%%%%%%%%%%%%%%%%%%%%%%%%%%%%%%%%%%%%%%%%%%%%%%%%%%%%%%%%%%%%%%%%
%%%%%%%%%%%%%%%%%%%%%%%%%%%%%%%%%%%%%%%%%%%%%%
\subsection{Quantum Otto Cycle}
This consists of four strokes; two quantum isochoric and two quantum adiabatic processes (see Fig.\ref{figQCC}). In the quantum adiabatic process, $B\to C$ and $D\to A$,  the occupational probability of the eigenstate is invariant, that is,  $\mathcal{P}_n(A)\!=\!\mathcal{P}_n(D)$ and $\mathcal{P}_n(B)\!=\!\mathcal{P}_n(C)$. This is because it satisfies the general adiabatic condition $d\mathcal{P}_n\!=\!0$. During this process, the system is isolated and does not exchange heat with the environment while the work is non-zero. 
\begin{figure}[!htbp]
\centering
\includegraphics[width=0.95 \columnwidth]{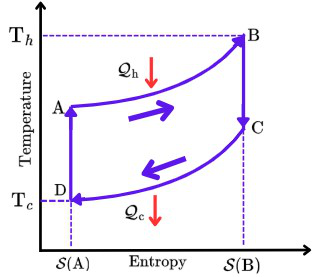} 
\caption{\justifying A pictorial illustration of an Otto machine The engine cycle consists of two adiabatic strokes ($B \to C$ and $D \to A$) where it is decoupled from the thermal baths and two isochoric strokes ($A \to B$ and $C \to D$) where the engine is coupled to two thermal baths at
temperatures $T_{\text{h}}$ and $T_{\text{c}}$, with $T_{\text{h}} > T_{\text{c}}$.}
\label{figQCC}  
\end{figure}
% using a $q$-deformed anharmonic oscillator as a working medium.

The quantum isochoric process represents $A\to B$ and $C\to D$. In this process, while heat is exchanged between the system and the reservoir, no work is done. Hence, the energy of the eigenvalue remains constant, that is, $\mathcal{E}^{\text{h}}\!=\!\mathcal{E} (B)\!=\!\mathcal{E}(A)$ and $\mathcal{E}^{\text{c}}\!=\!\mathcal{E}(D)\!=\!\mathcal{E}(C)$. Furthermore, in the isochoric process, $\mathcal{P}_n$ and $\mathcal{S}$ change until equilibrium is reached. Heat is released ($\mathcal{Q}_{\text{c}} <0$) when the entropy decreases in stroke $C\to D$, while heat is absorbed ($\mathcal{Q}_{\text{h}}<0$) when the entropy increases for $A \to B$, satisfying the Boltzmann distribution \cite{quan2007quantum}. The heat absorbed by the working medium during the heating process is
\begin{align}\label{QhotOtto}
    \mathcal{Q}_{\text{h}} = \int^{B}_{A}\, \mathcal{E}_n d\mathcal{P}_n =\,\sum_{n}\mathcal{E}_n^h[\mathcal{P}_n(B)-\mathcal{P}_n(D)],
\end{align}
where $\mathcal{E}^{\text{h}}$ is the $n^{th}$  energy of the system from $B\to C$. Furthermore, the heat released during the cooling process is 
\begin{align}\label{QcoldOtto}
    \mathcal{Q}_{\text{c}} = - \int^{D}_{C} \mathcal{E}_ndP_n \, = \sum\limits_{n}\mathcal{E}_n^c[\mathcal{P}_n(D)-\mathcal{P}_n(B)],
\end{align}
where $\mathcal{E}^{\text{c}}$ is the $n^\text{th}$  energy of the cooling process of the system from $D\to C$.
The  work done for the Otto heat engine  is the sum of  $\mathcal{Q}_{\text{h}} $ and $\mathcal{Q}_{\text{c}}$ given as \cite{quan2007quantum, Prakash2022};
\begin{equation}\label{WorkdoneOtto}
\mathcal{W}^{\text{QO}}  =\sum\limits_{n} \left[\mathcal{E}^{h}_{n}- \mathcal{E}^{\text{c}}_{n}\right]\left[\mathcal{P}_n(B)-\mathcal{P}_n(D)\right],
\end{equation}
and the corresponding efficiency of the $q$-deformed Morse oscillator quantum heat engine is \cite{quan2005quantum, kieu2006quantum}
\begin{equation}\label{EfficiencyOtto}
\eta_{otto} = \frac{\mathcal{W}^{\text{QO}}}{\mathcal{Q}_{\text{h}}}.
\end{equation} 
In what follows, we present the performance of three different scenarios for realizing a quantum Otto engine cycle using a diatomic molecule as the working medium.

\subsubsection{Changing width of the potential well}
Now consider the case in which the width of the potential well $\alpha/r_e$ and the temperature $T/r_{r_e}$ are varied during the cycle. Using the energy eigenvalue expression and Eq. \eqref{QcoldOtto}, we obtain the analytical expression for the input heat of the quantum Otto engine as,
\begin{eqnarray} 
     \mathcal{Q}^{\text{QO}}_{\text{h}}&=&   \frac{1}{2 \beta_\text{h} \sqrt{\pi }}\left(\frac{2 \gamma_{1}^{\text{h}} e^{\Lambda_{1}^{\text{h}}}}{\text{erfc}\left(\gamma_{1}^{\text{h}} \right)}+\sqrt{\pi } \left(-\frac{\beta_\text{h} \xi_\text{h}^2}{\beta_\text{c} \xi_\text{c}^2}+\Lambda_{3}^{\text{ch}}\right)\right) \nonumber\\ &+& \frac{1}{2  \sqrt{\pi }}\frac{\Lambda_{2}^{\text{c}} e^{ \Lambda_{1}^{\text{c}} }}{\sqrt{\beta_\text{c} \xi_\text{c}^2 (-p)} \text{erfc}\left( \gamma_{1}^{\text{c}} \right)},
\end{eqnarray}
where $\Lambda_{1}^{\text{c}}, \Lambda_{1}^{\text{h}},
\Lambda_{2}^{\text{ch}},
\Lambda_{3}^{\text{ch}}, \text{and} \,
2 \gamma_{1}^{\text{h}}$ are dimensionless parameters defined in Eq. \eqref{parameters1} of Appendix (\ref{Ottoequations}). Similarly, the expression for the output heat $\mathcal{Q}^{\text{QO}}_{\text{c}}$ reads
\begin{eqnarray}\label{Qcoldbody}
\mathcal{Q}^{\text{QO}}_{\text{c}} &=& \frac{1}{2 \sqrt{\pi }}\left(\frac{ 2 \gamma_{1}^{\text{c}} e^{ \Lambda_{1}^{\text{c}} }}{\beta_\text{c} \text{erfc}\left(  \gamma_{1}^{\text{c}} \right)}+\frac{ \Lambda_{2}^{\text{hc*}} e^{ \Lambda_{1}^{\text{h}} }}{\sqrt{ \beta _\text{h} \xi_\text{h}^2 (-p)} \text{erfc}\left( \gamma_{1}^{\text{h}} \right)} \right) \nonumber \\ &+& \frac{1}{2} \left(\frac{1}{\beta_\text{c}}+ \Lambda_{4}^{ch} \right),
\end{eqnarray}
where the dimensionless parameters 
$\Lambda_{2}^{\text{hc*}}, 2 \gamma_{1}^{\text{c}},
\gamma_{1}^{\text{h}}, \text{and}
\Lambda_{4}^{ch}$ are defined explicitly in Eq. \eqref{parameters1}, and \eqref{parameters2} of Appendix (\ref{Ottoequations}).
Using Eq. \eqref{WorkdoneOtto}, we obtain the work output as follows:
\begin{equation}
 \mathcal{W}^{\text{QO}} = \frac{ 1}{2 \sqrt{\pi }} \left( \frac{\Lambda_{10} +  \Lambda _{11} \sqrt{\beta_\text{h} \xi_\text{h}^2 (-p)} }{\sqrt{\beta_\text{h} \xi_\text{h}^2 (-p)}\text{$\beta $c} \beta_\text{h} \xi_\text{h}^2 \text{erfc}\left( \gamma_{1}^{\text{h}} \right)} -\frac{p \mathcal{R}_{0}}{\mathcal{R}_{1}} \right)
\end{equation}
where the dimensionless parameters $\Lambda_{5}^{\text{ch}}$, $\Lambda_{6}^{\text{ch}}$, $\Lambda_{7}^{\text{ch}}$, $\Lambda_{8}^{\text{ch}}$, $\Lambda_{9}^{\text{h}}$, $\Lambda_{9}^{\text{c}}$, $\Lambda_{10}, \Lambda_{11},  \Lambda_{12}$, $\mathcal{R}_{0}\!=\!\left(\Lambda_{12} +\Lambda_{13} \right)$, and $\mathcal{R}_{1}\!=\!\left(\beta_\text{c} \xi_\text{c}^2 (-p)\right)^{3/2} \text{erfc}\left( \gamma_{1}^{\text{c}} \right)$ are defined explicitly in Eqs. \eqref{parameters1}, \eqref{parameters2}  and  \eqref{parameters3} of Appendix (\ref{Ottoequations}). Finally, using Eq. \eqref{EfficiencyOtto} and \eqref{WorkdoneOttoapend}, an explicit analytical expression for efficiency is obtained (see Eq. \eqref{EfficiencyOttoappend} in Appendix (\ref{Ottoequations})). 
\begin{figure}[!htbp]
\centering
\includegraphics[width=0.98 \columnwidth]{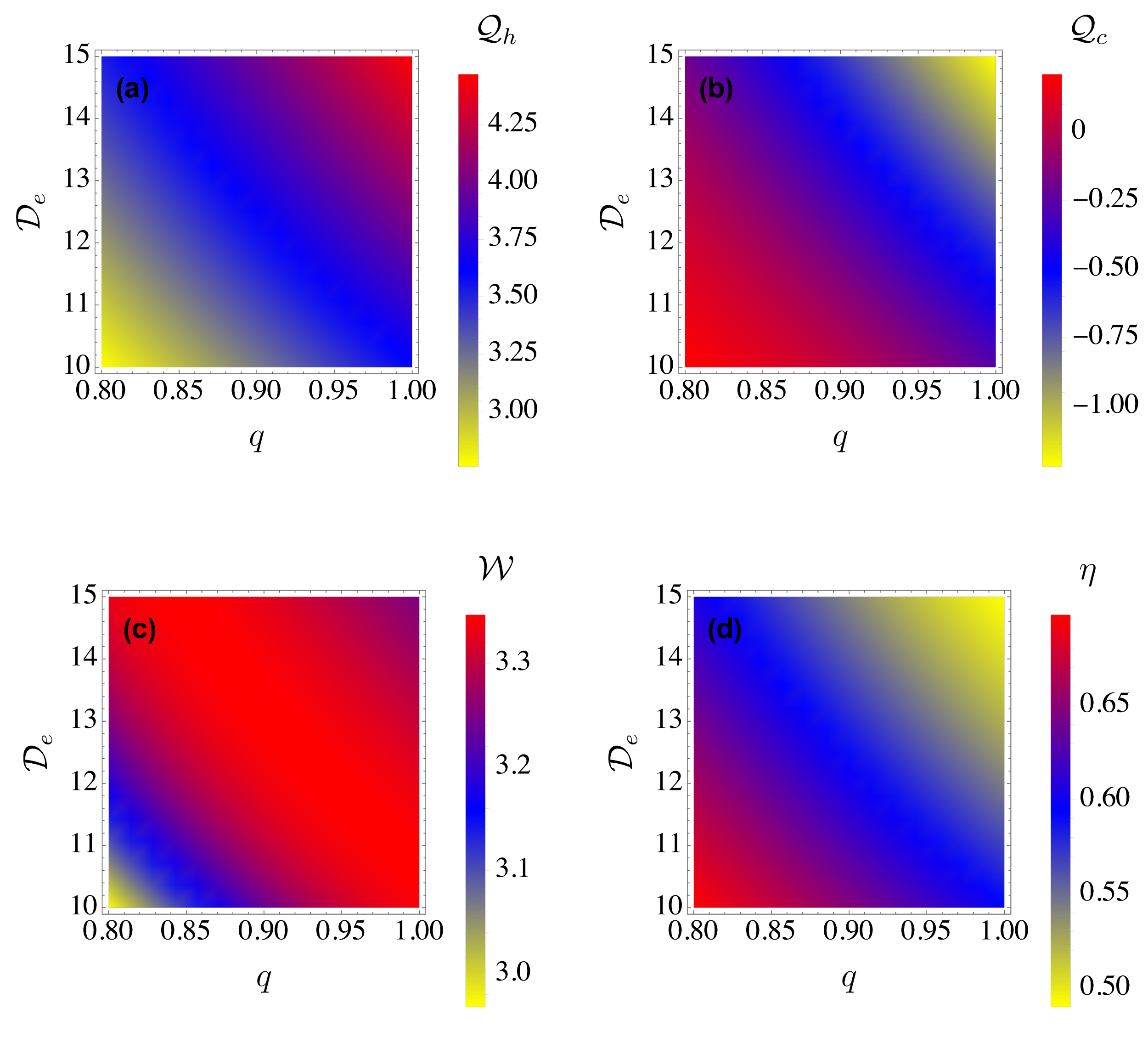} 
\caption{\justifying Quantum Otto engine with changing width of the potential well: (\textbf{a}) Heat exchange with the hot reservoir,  $\mathcal{Q}_{\text{h}}$,  (\textbf{b}) Heat exchange with the cold reservoir,  $\mathcal{Q}_{\text{c}}$  (\textbf{c}) Work output, $\mathcal{W}$, and  (\textbf{d}) Efficiency,  $\eta$  as a function of $\mathcal{D}_{e}$ and $q$. Parameters used: $\alpha_h\!=\!2.236 $, $\alpha_c\!=\!1$, $T_c\!=\!2$ and $T_h\!=\!10$.}
    \label{Fig3OttoCa}
\end{figure}
%%%%%%%%%%%%%%%%%%%%%%%%%%%%%%%%%%%%%%%%%%%%%%%%%%%%%%%%%%%%%%%%%%%%%%%%%%%%%%%%%%%%%%%%%%%%%%
We present the Otto engine thermodynamic figure of merit quantities in Fig. (\ref{Fig3OttoCa}). Figures \ref{Fig3OttoCa}(a) and (b) show that $\mathcal{Q}_h\!>\!0$ and $\mathcal{Q}_c\!<\!0$ are satisfied within a wide range of well width and potential deformation. In Fig. (\ref{Fig3OttoCa}) (c), we observe an increase in the work performed as both $\mathcal{D}_e/r_e$ and $q/r_e$ increase. This is particularly seen especially moving from left (low) $\mathcal{D}_e/r_e$ and $q/r_e$  towards the top-right region. The numerical result generally shows a positive value for the output work of the Otto heat engine. The value is very high in the region $0.95 r_e<q<0.99r_e$ and $13.9\, r_e\!<\!\mathcal{D}_e\!<\!15\,r_e$. Although a slight flip to a lower value is observed around $q\!=\!r_e$.  Based on the magnitudes of $\mathcal{Q}_{\text{h}}\!>\!0$ and $\mathcal{Q}_{\text{c}}\!<\!0$, we conclude that in this parameter regime considered, the system can operate as a quantum heat engine.  In Fig. \ref{Fig3OttoCa} (d), we present the efficiency of the quantum Otto heat engines as a function of $\mathcal{D}_e/r_e$ and $q/r_e$. We observe that for the parameter range considered, the efficiency of the Quantum Otto engine is lower than the efficiency of the Quantum Carnot Heat engine ($\eta_C\!=\!0.8$), which is in agreement with the literature \cite{de2021two, Prakash2022}. Focusing on Otto engine efficiency, we see that as $q\rightarrow1$, the efficiency decreases; this is in agreement with earlier studies by \citet{ozaydin2023powering} that focus on the $q$-deformed quantum oscillator as a working substance. Furthermore, this demonstrates that the deformation introduced is able to tune the efficiency and, by extension, other important figures of merit of our system. 

%%%%%%%%%%%%%%%%%%%%%%%%%%%%%%%%%%%%%%%%%%%%%%%%%%
\begin{figure}[!htbp]
\centering
\includegraphics[width=1 \columnwidth]{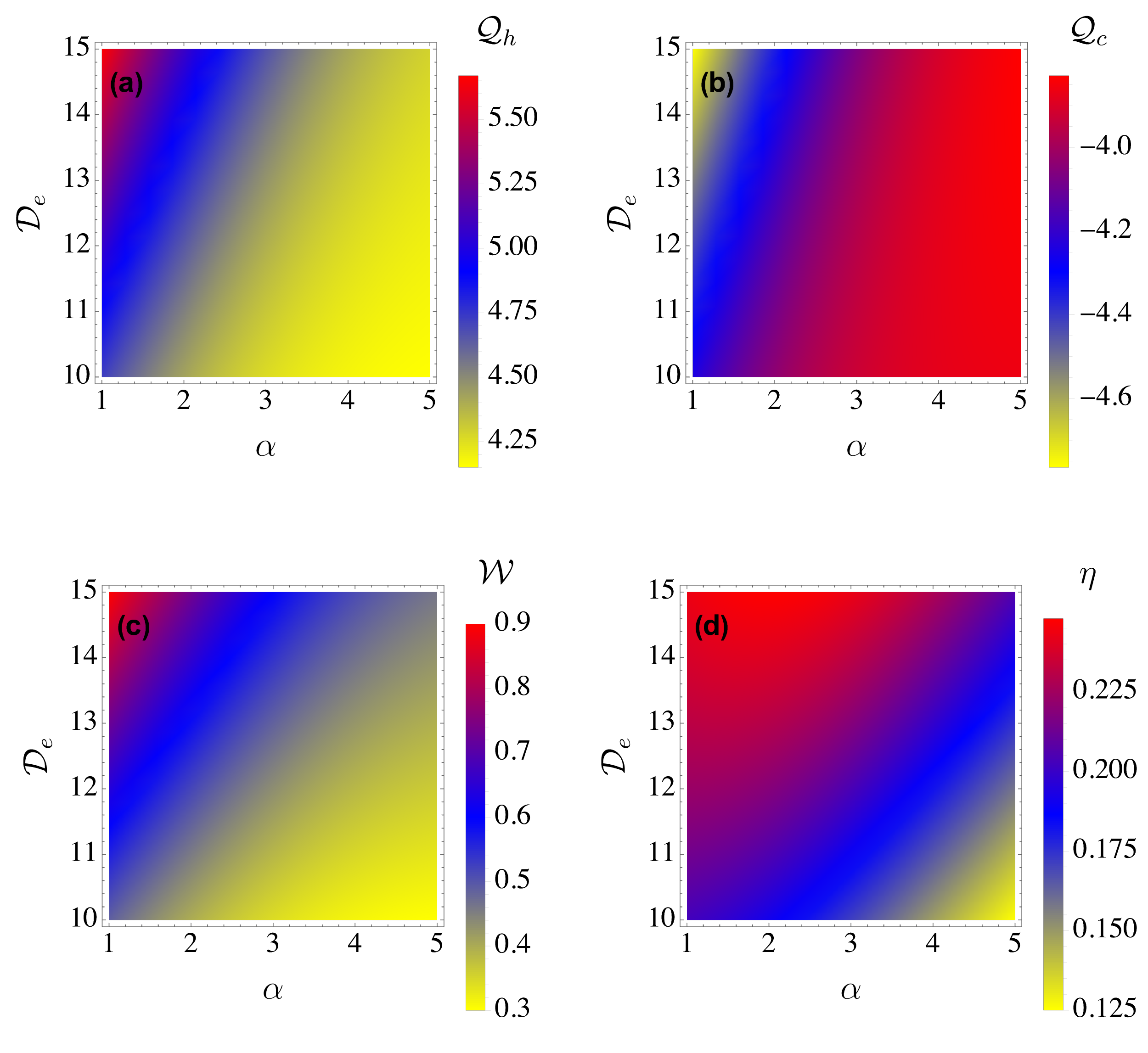} 
\caption{\justifying Quantum Otto cycle with the changing deformation parameter: (\textbf{a}) Heat exchange with the hot reservoir,  $\mathcal{Q}_{\text{h}}$ (\textbf{b}) Heat exchange  with the cold reservoir,  $\mathcal{Q}_{\text{c}}$  (\textbf{c}) Total work output, $\mathcal{W}$  (\textbf{d}) Efficiency  $\eta$  as function of $\mathcal{D}_{e}$ and $\alpha$. Parameters used: $q_c = 0.8 $, $q_h = 1$, $T_h = 10 $, and $T_c=2$. }
\label{fig4Otto}
\end{figure}

%%%%%%%%%%%%%%%%%%%%%%%%%%%%%%%%%%%%%%%%%%%%%%%%%%%%%%%%%%%%%%%%%%%%%%%%%%%%%%%%%%

\subsubsection{Changing the strength of the deformation parameter}
Here, the strength of the deformation parameter $(q/r_e)$ and the temperature $T/r_e$ vary during the cycle while the dissociation energy ($\mathcal{D}_e/r_e$) and the width of the potential well ($\alpha$)  are kept constant.  The work output, using Eq.\eqref{WorkdoneOtto}, is 
\begin{align} \label{WorkdoneOttoq}
\mathcal{W}^{\text{QO}}  =\sum\limits_{n} \left[\mathcal{E}^{\text{h}}_{n}(q_{\text{h}})- \mathcal{E}^{\text{c}}_{n} (q_\text{c})\right]\left[\mathcal{P}_n(B)-\mathcal{P}_n(\text{D})\right],
\end{align}
where $\mathcal{E}^{\text{h}}_{n}(q_\text{h})$, and $\mathcal{E}^{c}_{n}(q_\text{c})$ are the $n^\text{th}$ energy levels associated with
the two isochoric processes which imply constant energy levels. Also, for the quantum adiabatic strokes during
which the occupation probabilities of energy eigenstates remain unchanged. When the strength of the deformation parameter is varied, it is important to note that the Carnot reversibility condition is not satisfied \cite{quan2007quantum}. 

In Fig. (\ref{fig4Otto}), we show the  figure of merit of the performance of the quantum Otto engine using a diatomic molecule as its working substance for the case of varying deformation strength. Figures (\ref{fig4Otto}) (a) and (b) show the heat exchanges between the working substance and the heat reservoirs during the hot and cold isochoric process, respectively. Analyzing the magnitude of $\mathcal{Q}_{\text{h}}>0$ and $\mathcal{Q}_{\text{c}}<0$, as shown in Figs. (\ref{fig4Otto}) (a) and  (b),  we conclude that in this parameter regime considered, the system operates as a quantum heat engine. In Fig. (\ref{fig4Otto}) (c) show that the work output is high in the region of low $\alpha$ and high $\mathcal{D}_e/r_e$. This is particularly seen as we move from right (low) to left (high) $\mathcal{D}_e/r_e$ and $\alpha/r_e$  towards the top-region region. The plot generally shows a positive value of the work output for a wide range of parameters. This is also very intense in the region $r_e\!<\alpha<\!2r_e$ and  $13\, r_e <\mathcal{D}_e<15\, r_e$. 
In Fig. (\ref{fig4Otto}) (d), the efficiency is seen to be predominantly high in the region of low $\alpha/r_e$  and high $\mathcal{D}_e$.

\subsubsection{Changing the strength of the dissociation energy}
We consider the case of constructing a quantum Otto engine by varying the dissociation energy $(\mathcal{D}_e/r_e)$ and the temperature $T/r_e$. %In contrast, other parameters, such as the deformation parameter ($q/r_e$) and the width of the potential well ($\alpha/r_e$) will be constant during the cycle. 
The work performed in this case using \eqref{WorkdoneOtto} is 
\begin{equation} \label{WorkdoneOttoDe}
\mathcal{W}^{\text{QO}}  =\sum\limits_{n} \left[\mathcal{E}^{h}_{n}(\mathcal{D}_e^{\text{h}})- \mathcal{E}^{\text{c}}_{n} (\mathcal{D}_e^{\text{c}})\right]\left[\mathcal{P}_n(B)-\mathcal{P}_n(D)\right].
\end{equation}
Analyzing Figs. (\ref{fig5Otto}) (a) and (b) reveal that $\mathcal{Q}_{\text{h}}>0$ and $\mathcal{Q}_{\text{c}}<0$, over the range of parameters. The positive value of the total work output, Fig. (\ref{fig5Otto}) (c), implies that the system functions as a quantum heat engine. The work output is low in the high regime $\alpha/r_e$ and increases as $q/r_e$ increases. In Fig. (\ref{fig5Otto}) (d), the efficiency is observed to be predominantly high (low) in the region of high (low) $q/r_e$($\alpha$)  and vice versa.

%Again, in Fig. (\ref{fig5Otto}) (c), we see that the work output increases as the deformation parameter increases. The work output is low in the regime of high $\alpha/r_e$ and increases as low $q/r_e$ increases. Overall, the plot shows positive work. By analyzing the phase of  $\mathcal{Q}_{\text{h}}$ (a) and $\mathcal{Q}_{\text{c}}$ (b), it reveals $\mathcal{Q}_{\text{h}}>0$ and $\mathcal{Q}_{\text{c}}<0$, over the range of parameters considered, as shown in Figs. (\ref{fig5Otto}) (a) and  (b). Therefore,  we conclude that in this parameter regime considered, the system operates as a quantum heat engine. In Fig. (\ref{fig5Otto}) (d), the efficiency is also seen to be predominantly high (low) in the region of high(high) $q/r_e$($\alpha$)  and vice versa.
%%%%%%%%%%%%%%%%%%%%%%%%%%%%%%%%%%%%%%%%%%%%%%%%%%%%%%%%%%%%%%%%%%
\begin{figure}[!htbp]
\centering
\includegraphics[width=1 \columnwidth]{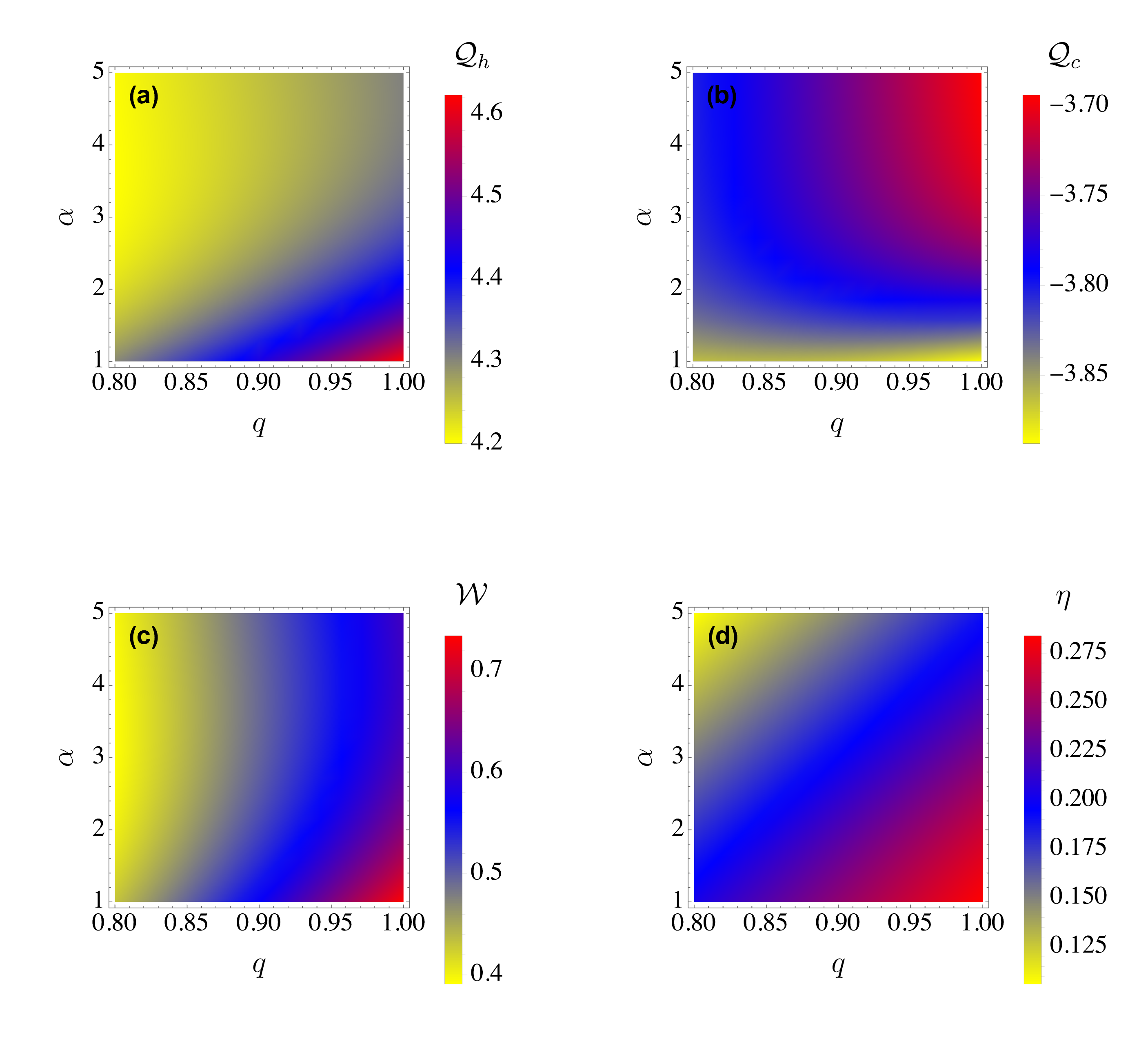} 
\caption{\justifying Quantum Otto cycle for varying dissociation energy strength: (\textbf{a}) Heat exchange with the hot reservoir,  $\mathcal{Q}_{\text{h}}$ vs $\alpha$ and $q$. (\textbf{b}) Heat exchange with the cold reservoir,  $\mathcal{Q}_{\text{c}}$ vs $\alpha$ and $q$. (\textbf{c})  Work output of Otto engine $\mathcal{W}$ vs $\alpha$ and $q$. (\textbf{d}) Efficiency  $\eta$  vs  $\alpha$ and $q$. Parameters used:$ \mathcal{D}_{e}^{\text{c}}=5$, $\mathcal{D}_{e}^{\text{h}}=10$, $T_h = 10 $, and $T_c=2$. }
\label{fig5Otto}
\end{figure}

\section{Conclusions}\label{conclusion}
We presented a theoretical proposal for a quantum thermal machine with the $q$-deformed Morse potential as a working medium. We discussed in detail the different regimes of operation for our heat engine. These modes of operation can be modulated by appropriate adjustment of the anharmonic potential parameters.  Furthermore, we present analytical expressions for the work done and the efficiency of the heat engine. We find that these figures of merits depend on the variations of the potential parameters. It is also noted that the performance of both the engine can be regulated or controlled by adjusting the deformation parameter and other system parameters.
The potential platform for the realization of our model is a laser-cooled trapped ion as a microscopic heat machine, proposed by \citet{gelbwaser2018single}. However, in our context, the trap will be anharmonic, which will be more suitable for trapping a diatomic molecule. The experimental considerations for the realization of a tunable anharmonic trap have been presented by \citet{home2011normal}. Moreover, the deformation parameter $q$ can be interpreted as an effective control parameter that encodes deviations from the standard Morse potential, which can be engineered. Our findings may inspire several developments in the emerging field of quantum technologies, such as the design of quantum thermal machines based on anharmonic models. 

\section*{Acknowledgements}
COE and NA  would also like to acknowledge the Universiti Malaysia Perlis (Funding No. 9004-00100 Special Research Grant-International Postdoctoral) for funding this project. COE and EPI gratefully acknowledge the support provided by the Tertiary Education Trust Fund-Institutional Based Research (TETFUND-IBR) grant scheme with grant number: NOUN/DRA/TETFUNDAW/VOLI. 
%COE and NA acknowledge support from the Long Term Research Grant Scheme (LRGS) with grant number LRGS/1/2020/UM/01/5/2 (9012-00009), provided by the Ministry of Higher Education of Malaysia (MOHE). 
%OA acknowledges the Newcastle University Academic Track Fellowship.
%N. Ali and Inyang, E.P. acknowledge the support from the UniMAP Special Research Grant for International Postdoctoral Fellowship with grant number: 9004-00100
\appendix
\section{Solutions of the Schrödinger Equation for the q-deformed Morse Potential}\label{Solutionsappend}
The Schrodinger equation in one-dimensional has the form

    \begin{equation} \label{SchrodingerEquation}
\left(-\frac{\hbar^2}{2\mu r_e^2} \frac{d^2}{dx^2} + \mathcal{V}_q(x) \right) \psi_{n}(x) = \mathcal{E}_{n} \psi_{n}(x)
\end{equation}
Considering the q-deformed Morse potential as
\begin{equation}\label{qdeformedMorse}
\mathcal{V}_q(x) = \mathcal{D}_e(e^{-2 \xi  x}-2  q e^{-\xi x})    
\end{equation}
Substituting Eq. \eqref{qdeformedMorse} in  Eq. \eqref{SchrodingerEquation}, we obtain
\begin{equation} \label{decomposedSE}
\left( \frac{d^2}{dx^2}-\frac{2 \mu \mathcal{D}_e}{\hbar^2} e^{-2 \xi  x}+\frac{4 \mu q \mathcal{D}_e}{\hbar^2}e^{-\xi x}  + \frac{2 \mu}{\hbar^2} \mathcal{E}_{n} \right) \psi_{n}(x) =  0
\end{equation}
Assuming $y = \delta e^{-\xi x}$ , Eq. \eqref{decomposedSE} takes the form
\begin{equation}\label{diff1}
\left( \frac{d^2}{dy^2} +\frac{1}{y}\frac{d}{dy}+\frac{ -\text{s}^2 }{y^2}+ \frac{\delta q}{2y} -\frac{1}{4}\right) \psi_{n}(y) =  0
\end{equation}
where $s= \sqrt{-\frac{2 \mu r_e^2\mathcal{E}_n}{\hbar^2\xi^2}}$, $\delta= \sqrt{\frac{8 \mu r_e^2  \mathcal{D}_e}{\hbar^2\xi^2}}$
Looking at the behaviour of the wave function at the origin and at infinity, we can take the following ansatz as
\begin{equation}\label{ansatz}
    \psi (y) = \text{e}^{-\frac{y}{2}} y^{s} \mathcal{F}(y)
\end{equation}
Substituting the wave function ansatz \eqref{ansatz} into Eq. \eqref{diff1}, this yields
\begin{equation}
y \frac{d^2}{dy^2} \mathcal{F}(y) +(2s+1-y) \frac{d}{dy}\mathcal{F}(y) -\left(s+\frac{1-\delta q}{2} \right) \mathcal{F}(y) =0,
\end{equation}
whose solutions are the confluent hypergeometric functions as
\begin{equation}
 \mathcal{F}(y) =\mathcal{N} \mathcal{F} \left(s+\frac{1-\delta q}{2}, 2s+1;y \right)   
\end{equation}
Thus, we obtain the quantum condition
\begin{equation}
s+\frac{1-\delta q}{2}=-n,    n=0,1,2...,    
\end{equation}
From which we obtain the energy as
\begin{equation}\label{energyequation}
    \mathcal{E}_{n}= -\frac{\hbar^2\alpha^2}{2 \mu} \left(q \sqrt{\frac{2 \mu \mathcal{D}_e}{\hbar^2\alpha^2}}  -\left(n+\frac{1}{2}\right)       \right)^2, 
\end{equation}
where $n=0,1,2,3,..., n_{\text{max}}$, and  $n_{\text{max}}$ can be derived by $\frac{d\mathcal{E}_{n}}{dn}=0$, thus we obtain: $n_{\text{max}}\leq (\lambda q-1/2) $.
%\begin{equation}
%n_{\text{max}}\leq (\lambda q-1/2) 
%\end{equation}
For completeness, the wave function of the q-deformed Morse potential can be written as:
\begin{align}
\psi (y) = \mathcal{N}_{n}^{\delta} \text{e}^{-\frac{y}{2}} y^{s} L_{n}^{2s}(y),
%{}_1\mathcal{F}_{1} \left(s+\frac{1-\delta q}{2}, 2s+1;y \right)    
\end{align}
where $ L_{n}^{2s}(y)$ are the associated Laguerre functions and  $\mathcal{N}_{n}^{\delta}$ is the normalization constant.

\section{Quantum Otto Cycle: \texorpdfstring{$\mathcal{Q}^{\text{QO}}_{\text{h}}$, $\mathcal{Q}^{\text{QO}}_{\text{c}}$ and $\mathcal{W}^{\text{QO}}$}{Quantum Otto Cycle: Qh, Qc and W}} \label{Ottoequations}
Here, we present the expressions of the thermodynamic quantities used for the Otto engine performance analysis. Employing the energy equation and Eq. \eqref{QhotOtto}, we obtain the analytical expression for $\mathcal{Q}^{\text{QO}}_{\text{h}}$ as follows:
%\begin{equation} \label{Qhotapend}
%     \mathcal{Q}^{\text{QO}}_{\text{h}}=   \frac{\frac{\frac{ 2 \gamma_{1}^{\text{h}} e^{\Lambda_{1}^{\text{h}}}}{\text{erfc}\left(\gamma_{1}^{\text{h}} \right)}+\sqrt{\pi } \left(-\frac{\beta_\text{h} \xi_\text{h}^2}{\beta_\text{c} \xi_\text{c}^2}+\Lambda_{3}^{\text{ch}}\right)}{\beta_\text{h}}+\frac{\Lambda_{2}^{\text{c}} e^{ \Lambda_{1}^{\text{c}} }}{\sqrt{\beta_\text{c} \xi_\text{c}^2 (-p)} \text{erfc}\left( \gamma_{1}^{\text{c}} \right)}}{2 \sqrt{\pi }}
%\end{equation}
\begin{equation} \label{Qhotapend}
\mathcal{Q}^{\text{QO}}_{\text{h}}=   \frac{  \gamma_{1}^{\text{h}} e^{\Lambda_{1}^{\text{h}}}}{ \sqrt{\pi} \tau^h}+\frac{\Lambda_{3}^{\text{ch}}}{2\beta_{\text{h}}}-\frac{ \xi_\text{h}^2}{2 \beta_{\text{c}}\xi_\text{c}^2}+\frac{\Lambda_{2}^{\text{c}} e^{ \Lambda_{1}^{\text{c}} }}{2\sqrt{\pi\beta_{\text{c}}\xi_\text{c}^2 (-p)\tau^c} }
\end{equation}

where 
\begin{eqnarray} \label{parameters1}
\tau^{c(h)}=\beta_{\text{c(h)}} \operatorname{erfc}\left( \gamma_{1}^{\text{c(h)}} \right),\\
\Lambda_{1}^{\text{c}}=\frac{1}{4} \beta_\text{c} \xi_\text{c}^2 p (1-2 \lambda_\text{c} q)^2,\\
\Lambda_{1}^{\text{h}}=\frac{1}{4} \beta_\text{h} \xi_\text{h}^2 p (1-2 \lambda_\text{h} q)^2,\\
\Lambda_{2}^{\text{ch}} = \xi_\text{h}^2 p (2 q (\lambda_\text{c}-2 \lambda_\text{h})+1),\\
\Lambda_{3}^{\text{ch}}=2 \beta_\text{h} \xi_\text{h}^2 p q^2 (\lambda_\text{c}-\lambda_\text{h})^2+1,\\
2 \gamma_{1}^{\text{h}}= (1-2 \lambda_\text{h} q) \sqrt{\beta_\text{h} \xi_\text{h}^2 (-p)}\\
\end{eqnarray}
Similarly, employing the energy equation and Eq. \eqref{QcoldOtto}, we obtain the analytical expression for $\mathcal{Q}^{\text{QO}}_{\text{c}}$ as follows:
\begin{eqnarray}\label{Qcoldapend}
\mathcal{Q}^{\text{QO}}_{\text{c}}&=& \frac{ \gamma_{1}^{\text{c}} e^{ \Lambda_{1}^{\text{c}} }}{\sqrt{\pi}\tau^{c}} + \frac{\Lambda_{2}^{\text{hc*}} e^{ \Lambda_{1}^{\text{h}} }}{2 \sqrt{\pi}\sqrt{ \beta_\text{h} \xi_\text{h}^2 (-p)} \text{erfc}\left(\gamma_{1}^{\text{h}}\right)}\nonumber\\ &+& \left(\frac{1}{2\beta_\text{c}}+ \frac{\Lambda_{4}^{ch}}{2} \right)
\end{eqnarray}
where 
 \begin{eqnarray}\label{parameters2}
 \Lambda_{2}^{\text{hc*}}= \xi_\text{c}^2 p (2 q (\lambda_\text{h}-2 \lambda_\text{c})+1),\\
2 \gamma_{1}^{\text{c}} = (1-2 \text{$\lambda $c} q) \sqrt{\beta_\text{c} \xi_\text{c}^2 (-p)}\\
\gamma_{1}^{\text{h}}=\frac{1}{2} (1-2 \lambda_\text{h} q) \sqrt{\beta_\text{h} \xi_\text{h}^2 (-p)}\\
\Lambda_{4}^{ch}=\xi_\text{c}^2 \left(2 p q^2 (\lambda_\text{c}-\lambda_\text{h})^2-\frac{1}{\beta_\text{h} \xi_\text{h}^2}\right)
\end{eqnarray}
%\begin{widetext}
Similarly, using Eq. \eqref{WorkdoneOtto}, we obtain the work output as follows:
\begin{equation}\label{WorkdoneOttoapend}
\mathcal{W}^{\text{QO}} = \frac{1}{2 \sqrt{\pi }}\left(\frac{\eta+ \Lambda _{11} }{\tau^h \beta_\text{c}  \xi_\text{h}^2}-\frac{p \left(\Lambda_{12} +\Lambda_{13} \right)}{\left( \xi_\text{c}^2 (-p)\right)^{3/2} \tau^c}\right)
\end{equation}
where
 \begin{align}\label{parameters3}
 \eta =\frac{\Lambda_{10}}{\sqrt{\beta_\text{h} \xi_\text{h}^2 (-p)}},\\
\Lambda_{5}^{\text{ch}}=  2 \beta_\text{c} \xi_\text{c}^2 p q^2 (\lambda_\text{c}-\lambda_\text{h})^2-1,\\
\Lambda_{6}^{\text{ch}}= \xi_\text{h}^2 (2 \lambda_\text{c} q-4 \lambda_\text{h} q+1)+\xi_\text{c}^2 (2 \lambda_\text{c} q-1)\\
\Lambda_{7}^{\text{ch}}=\xi_\text{c}^2 (-4 \lambda_\text{c} q+2 \lambda_\text{h} q+1)+\xi_\text{h}^2 (2 \lambda_\text{h} q-1)\\
\Lambda_{8}^{\text{ch}}= \xi_\text{h}^2 \left(\beta_\text{c}+\beta_\text{h}+2 \beta_\text{c} \beta_\text{h} \xi_\text{c}^2 p q^2 (\lambda_\text{c}-\lambda_\text{h})^2\right)-\beta_\text{c} \xi_\text{c}^2\\ 
\Lambda_{9}^{\text{h}} =\frac{1}{2} \sqrt{ \beta_\text{h}} \xi_\text{h} \sqrt{p} (1-2 \lambda_\text{h} q)\\
\Lambda_{9}^{\text{c}} =\frac{1}{2} \sqrt{\beta_\text{c}} \xi_\text{c} \sqrt{p} (1-2 \lambda_\text{c} q)\\
\Lambda_{10}=\sqrt{\pi } \sqrt{\beta_\text{h}} \xi_\text{h} \sqrt{p}\, \text{erfi}\left( \Lambda_{9}^{\text{h}} \right) \left( \Lambda_{8}^{\text{ch}}\right)\\
\Lambda_{11}= \sqrt{\pi } \left( \Lambda_{8}^{\text{ch}}\right)-\beta_\text{c} \sqrt{\beta_\text{h} \xi_\text{h}^2 (-p)} e^{ \Lambda_{1}^{\text{h}} } \left(\Lambda_{7}^{\text{ch}}  \right)\\
\Lambda_{12}=\sqrt{\pi } \sqrt{\beta_\text{c}} \xi_\text{c} \xi_\text{h}^2 \sqrt{p} \, \text{erfi} \left(  \Lambda_{9}^{\text{c}} \right) \left( \Lambda_{5}^{\text{ch}} \right)\\
\Lambda_{13}=\sqrt{\pi } \xi_\text{h}^2 \sqrt{\beta_\text{c} \xi_\text{c}^2 (-p)} \left( \Lambda_{5}^{\text{ch}}  \right)+\beta_\text{c} \xi_\text{c}^2 p\, e^{\Lambda_{1}^{\text{c}} } \left( \Lambda_{6}^{\text{ch}}  \right)
\end{align}
Finally, using Eq. \eqref{EfficiencyOtto} and \eqref{WorkdoneOttoapend}, we obtain an explicit analytical expression for the  efficiency as follows:
\begin{equation}\label{EfficiencyOttoappend}
\eta^{{\text{QO}}} =\frac{ \frac{ \eta+ \Lambda _{11} }{ \beta_\text{c} \xi_\text{h}^2 \tau^h}-\frac{p \left(\Lambda_{12} +\Lambda_{13} \right)}{(\beta_\text{c} \xi_\text{c}^2 (-p))^{3/2} \text{erfc}\left( \gamma_{1}^{\text{c}} \right)}}{ \frac{ 2 \gamma_{1}^{\text{h}} e^{\Lambda_{1}^{\text{h}}}}{\tau^h}+\sqrt{\pi } \left(-\frac{ \xi_\text{h}^2}{\beta_\text{c} \xi_\text{c}^2}+\frac{\Lambda_{3}^{\text{ch}}}{\beta_\text{h}}\right)+\frac{\Lambda_{2}^{\text{c}} e^{ \Lambda_{1}^{\text{c}} }}{\sqrt{\beta_\text{c} \xi_\text{c}^2 (-p)} \text{erfc}\left( \gamma_{1}^{\text{c}} \right)}} 
\end{equation}

\bibliography{References}

\begin{thebibliography}{64}%
\makeatletter
\providecommand \@ifxundefined [1]{%
 \@ifx{#1\undefined}
}%
\providecommand \@ifnum [1]{%
 \ifnum #1\expandafter \@firstoftwo
 \else \expandafter \@secondoftwo
 \fi
}%
\providecommand \@ifx [1]{%
 \ifx #1\expandafter \@firstoftwo
 \else \expandafter \@secondoftwo
 \fi
}%
\providecommand \natexlab [1]{#1}%
\providecommand \enquote  [1]{``#1''}%
\providecommand \bibnamefont  [1]{#1}%
\providecommand \bibfnamefont [1]{#1}%
\providecommand \citenamefont [1]{#1}%
\providecommand \href@noop [0]{\@secondoftwo}%
\providecommand \href [0]{\begingroup \@sanitize@url \@href}%
\providecommand \@href[1]{\@@startlink{#1}\@@href}%
\providecommand \@@href[1]{\endgroup#1\@@endlink}%
\providecommand \@sanitize@url [0]{\catcode `\\12\catcode `\$12\catcode `\&12\catcode `\#12\catcode `\^12\catcode `\_12\catcode `\%12\relax}%
\providecommand \@@startlink[1]{}%
\providecommand \@@endlink[0]{}%
\providecommand \url  [0]{\begingroup\@sanitize@url \@url }%
\providecommand \@url [1]{\endgroup\@href {#1}{\urlprefix }}%
\providecommand \urlprefix  [0]{URL }%
\providecommand \Eprint [0]{\href }%
\providecommand \doibase [0]{https://doi.org/}%
\providecommand \selectlanguage [0]{\@gobble}%
\providecommand \bibinfo  [0]{\@secondoftwo}%
\providecommand \bibfield  [0]{\@secondoftwo}%
\providecommand \translation [1]{[#1]}%
\providecommand \BibitemOpen [0]{}%
\providecommand \bibitemStop [0]{}%
\providecommand \bibitemNoStop [0]{.\EOS\space}%
\providecommand \EOS [0]{\spacefactor3000\relax}%
\providecommand \BibitemShut  [1]{\csname bibitem#1\endcsname}%
\let\auto@bib@innerbib\@empty
%</preamble>
\bibitem [{\citenamefont {Cengel}(2011)}]{cengel2011thermodynamics}%
  \BibitemOpen
  \bibfield  {author} {\bibinfo {author} {\bibfnamefont {Y.~A.}\ \bibnamefont {Cengel}},\ }\href@noop {} {\bibinfo {title} {Thermodynamics: an engineering approach}} (\bibinfo {year} {2011})\BibitemShut {NoStop}%
\bibitem [{\citenamefont {Scovil}\ and\ \citenamefont {Schulz-DuBois}(1959)}]{scovil1959three}%
  \BibitemOpen
  \bibfield  {author} {\bibinfo {author} {\bibfnamefont {H.~E.}\ \bibnamefont {Scovil}}\ and\ \bibinfo {author} {\bibfnamefont {E.~O.}\ \bibnamefont {Schulz-DuBois}},\ }\bibfield  {title} {\bibinfo {title} {Three-level masers as heat engines},\ }\href {https://doi.org/10.1103/PhysRevLett.2.262} {\bibfield  {journal} {\bibinfo  {journal} {Physical Review Letters}\ }\textbf {\bibinfo {volume} {2}},\ \bibinfo {pages} {262} (\bibinfo {year} {1959})}\BibitemShut {NoStop}%
\bibitem [{\citenamefont {Myers}\ \emph {et~al.}(2022)\citenamefont {Myers}, \citenamefont {Abah},\ and\ \citenamefont {Deffner}}]{myers2022quantum}%
  \BibitemOpen
  \bibfield  {author} {\bibinfo {author} {\bibfnamefont {N.~M.}\ \bibnamefont {Myers}}, \bibinfo {author} {\bibfnamefont {O.}~\bibnamefont {Abah}},\ and\ \bibinfo {author} {\bibfnamefont {S.}~\bibnamefont {Deffner}},\ }\bibfield  {title} {\bibinfo {title} {Quantum thermodynamic devices: From theoretical proposals to experimental reality},\ }\href {https://doi.org/10.1116/5.0083192} {\bibfield  {journal} {\bibinfo  {journal} {AVS Quantum Science}\ }\textbf {\bibinfo {volume} {4}} (\bibinfo {year} {2022})}\BibitemShut {NoStop}%
\bibitem [{\citenamefont {Quan}\ \emph {et~al.}(2007)\citenamefont {Quan}, \citenamefont {Liu}, \citenamefont {Sun},\ and\ \citenamefont {Nori}}]{quan2007quantum}%
  \BibitemOpen
  \bibfield  {author} {\bibinfo {author} {\bibfnamefont {H.-T.}\ \bibnamefont {Quan}}, \bibinfo {author} {\bibfnamefont {Y.-x.}\ \bibnamefont {Liu}}, \bibinfo {author} {\bibfnamefont {C.-P.}\ \bibnamefont {Sun}},\ and\ \bibinfo {author} {\bibfnamefont {F.}~\bibnamefont {Nori}},\ }\bibfield  {title} {\bibinfo {title} {Quantum thermodynamic cycles and quantum heat engines},\ }\href {https://doi.org/10.1103/PhysRevE.76.031105} {\bibfield  {journal} {\bibinfo  {journal} {Physical Review E}\ }\textbf {\bibinfo {volume} {76}},\ \bibinfo {pages} {031105} (\bibinfo {year} {2007})}\BibitemShut {NoStop}%
\bibitem [{\citenamefont {Ryan}\ \emph {et~al.}(2008)\citenamefont {Ryan}, \citenamefont {Moussa}, \citenamefont {Baugh},\ and\ \citenamefont {Laflamme}}]{ryan2008spin}%
  \BibitemOpen
  \bibfield  {author} {\bibinfo {author} {\bibfnamefont {C.}~\bibnamefont {Ryan}}, \bibinfo {author} {\bibfnamefont {O.}~\bibnamefont {Moussa}}, \bibinfo {author} {\bibfnamefont {J.}~\bibnamefont {Baugh}},\ and\ \bibinfo {author} {\bibfnamefont {R.}~\bibnamefont {Laflamme}},\ }\bibfield  {title} {\bibinfo {title} {Spin based heat engine: demonstration of multiple rounds of algorithmic cooling},\ }\href {https://doi.org/10.1103/PhysRevLett.100.140501} {\bibfield  {journal} {\bibinfo  {journal} {Physical Review Letters}\ }\textbf {\bibinfo {volume} {100}},\ \bibinfo {pages} {140501} (\bibinfo {year} {2008})}\BibitemShut {NoStop}%
\bibitem [{\citenamefont {Friedenberger}\ and\ \citenamefont {Lutz}(2017)}]{friedenberger2017quantum}%
  \BibitemOpen
  \bibfield  {author} {\bibinfo {author} {\bibfnamefont {A.}~\bibnamefont {Friedenberger}}\ and\ \bibinfo {author} {\bibfnamefont {E.}~\bibnamefont {Lutz}},\ }\bibfield  {title} {\bibinfo {title} {When is a quantum heat engine quantum?},\ }\href {https://doi.org/10.1209/0295-5075/120/10002} {\bibfield  {journal} {\bibinfo  {journal} {Europhysics Letters}\ }\textbf {\bibinfo {volume} {120}},\ \bibinfo {pages} {10002} (\bibinfo {year} {2017})}\BibitemShut {NoStop}%
\bibitem [{\citenamefont {Das}\ and\ \citenamefont {Ghosh}(2019)}]{das2019measurement}%
  \BibitemOpen
  \bibfield  {author} {\bibinfo {author} {\bibfnamefont {A.}~\bibnamefont {Das}}\ and\ \bibinfo {author} {\bibfnamefont {S.}~\bibnamefont {Ghosh}},\ }\bibfield  {title} {\bibinfo {title} {Measurement based quantum heat engine with coupled working medium},\ }\href {https://doi.org/10.3390/e21111131} {\bibfield  {journal} {\bibinfo  {journal} {Entropy}\ }\textbf {\bibinfo {volume} {21}},\ \bibinfo {pages} {1131} (\bibinfo {year} {2019})}\BibitemShut {NoStop}%
\bibitem [{\citenamefont {Huang}\ \emph {et~al.}(2014)\citenamefont {Huang}, \citenamefont {Niu}, \citenamefont {Xiu},\ and\ \citenamefont {Yi}}]{huang2014quantum}%
  \BibitemOpen
  \bibfield  {author} {\bibinfo {author} {\bibfnamefont {X.-L.}\ \bibnamefont {Huang}}, \bibinfo {author} {\bibfnamefont {X.-Y.}\ \bibnamefont {Niu}}, \bibinfo {author} {\bibfnamefont {X.-M.}\ \bibnamefont {Xiu}},\ and\ \bibinfo {author} {\bibfnamefont {X.-X.}\ \bibnamefont {Yi}},\ }\bibfield  {title} {\bibinfo {title} {Quantum stirling heat engine and refrigerator with single and coupled spin systems},\ }\href {https://doi.org/10.1140/epjd/e2013-40536-0} {\bibfield  {journal} {\bibinfo  {journal} {The European Physical Journal D}\ }\textbf {\bibinfo {volume} {68}},\ \bibinfo {pages} {1} (\bibinfo {year} {2014})}\BibitemShut {NoStop}%
\bibitem [{\citenamefont {Fadaie}\ \emph {et~al.}(2018)\citenamefont {Fadaie}, \citenamefont {Yunt},\ and\ \citenamefont {M{\"u}stecapl{\i}o{\u{g}}lu}}]{fadaie2018topological}%
  \BibitemOpen
  \bibfield  {author} {\bibinfo {author} {\bibfnamefont {M.}~\bibnamefont {Fadaie}}, \bibinfo {author} {\bibfnamefont {E.}~\bibnamefont {Yunt}},\ and\ \bibinfo {author} {\bibfnamefont {{\"O}.~E.}\ \bibnamefont {M{\"u}stecapl{\i}o{\u{g}}lu}},\ }\bibfield  {title} {\bibinfo {title} {Topological phase transition in quantum-heat-engine cycles},\ }\href {https://doi.org/https://doi.org/10.1103/PhysRevE.98.052124} {\bibfield  {journal} {\bibinfo  {journal} {Physical Review E}\ }\textbf {\bibinfo {volume} {98}},\ \bibinfo {pages} {052124} (\bibinfo {year} {2018})}\BibitemShut {NoStop}%
\bibitem [{\citenamefont {Mani}\ and\ \citenamefont {Benjamin}(2017)}]{mani2017strained}%
  \BibitemOpen
  \bibfield  {author} {\bibinfo {author} {\bibfnamefont {A.}~\bibnamefont {Mani}}\ and\ \bibinfo {author} {\bibfnamefont {C.}~\bibnamefont {Benjamin}},\ }\bibfield  {title} {\bibinfo {title} {Strained-graphene-based highly efficient quantum heat engine operating at maximum power},\ }\href {https://doi.org/10.1103/PhysRevE.96.032118} {\bibfield  {journal} {\bibinfo  {journal} {Physical Review E}\ }\textbf {\bibinfo {volume} {96}},\ \bibinfo {pages} {032118} (\bibinfo {year} {2017})}\BibitemShut {NoStop}%
\bibitem [{\citenamefont {Mu\~noz}\ and\ \citenamefont {Pe\~na}(2012)}]{munoz2012quantum}%
  \BibitemOpen
  \bibfield  {author} {\bibinfo {author} {\bibfnamefont {E.}~\bibnamefont {Mu\~noz}}\ and\ \bibinfo {author} {\bibfnamefont {F.~J.}\ \bibnamefont {Pe\~na}},\ }\bibfield  {title} {\bibinfo {title} {Quantum heat engine in the relativistic limit: The case of a dirac particle},\ }\href {https://link.aps.org/doi/10.1103/PhysRevE.86.061108} {\bibfield  {journal} {\bibinfo  {journal} {Phys. Rev. E}\ }\textbf {\bibinfo {volume} {86}},\ \bibinfo {pages} {061108} (\bibinfo {year} {2012})}\BibitemShut {NoStop}%
\bibitem [{\citenamefont {Pe{\~n}a}\ \emph {et~al.}(2016)\citenamefont {Pe{\~n}a}, \citenamefont {Ferr{\'e}}, \citenamefont {Orellana}, \citenamefont {Rojas},\ and\ \citenamefont {Vargas}}]{pena2016optimization}%
  \BibitemOpen
  \bibfield  {author} {\bibinfo {author} {\bibfnamefont {F.~J.}\ \bibnamefont {Pe{\~n}a}}, \bibinfo {author} {\bibfnamefont {M.}~\bibnamefont {Ferr{\'e}}}, \bibinfo {author} {\bibfnamefont {P.}~\bibnamefont {Orellana}}, \bibinfo {author} {\bibfnamefont {R.~G.}\ \bibnamefont {Rojas}},\ and\ \bibinfo {author} {\bibfnamefont {P.}~\bibnamefont {Vargas}},\ }\bibfield  {title} {\bibinfo {title} {Optimization of a relativistic quantum mechanical engine},\ }\href {https://doi.org/10.1103/PhysRevE.94.022109} {\bibfield  {journal} {\bibinfo  {journal} {Physical Review E}\ }\textbf {\bibinfo {volume} {94}},\ \bibinfo {pages} {022109} (\bibinfo {year} {2016})}\BibitemShut {NoStop}%
\bibitem [{\citenamefont {Kieu}(2006)}]{kieu2006quantum}%
  \BibitemOpen
  \bibfield  {author} {\bibinfo {author} {\bibfnamefont {T.~D.}\ \bibnamefont {Kieu}},\ }\bibfield  {title} {\bibinfo {title} {Quantum heat engines, the second law and maxwell's daemon},\ }\href {https://doi.org/https://doi.org/10.1140/epjd/e2006-00075-5} {\bibfield  {journal} {\bibinfo  {journal} {The European Physical Journal D-Atomic, Molecular, Optical and Plasma Physics}\ }\textbf {\bibinfo {volume} {39}},\ \bibinfo {pages} {115} (\bibinfo {year} {2006})}\BibitemShut {NoStop}%
\bibitem [{\citenamefont {Quan}\ \emph {et~al.}(2005{\natexlab{a}})\citenamefont {Quan}, \citenamefont {Zhang},\ and\ \citenamefont {Sun}}]{quan2005quantummulti}%
  \BibitemOpen
  \bibfield  {author} {\bibinfo {author} {\bibfnamefont {H.}~\bibnamefont {Quan}}, \bibinfo {author} {\bibfnamefont {P.}~\bibnamefont {Zhang}},\ and\ \bibinfo {author} {\bibfnamefont {C.}~\bibnamefont {Sun}},\ }\bibfield  {title} {\bibinfo {title} {Quantum heat engine with multilevel quantum systems},\ }\href {https://doi.org/10.1103/PhysRevE.72.056110} {\bibfield  {journal} {\bibinfo  {journal} {Physical Review E—Statistical, Nonlinear, and Soft Matter Physics}\ }\textbf {\bibinfo {volume} {72}},\ \bibinfo {pages} {056110} (\bibinfo {year} {2005}{\natexlab{a}})}\BibitemShut {NoStop}%
\bibitem [{\citenamefont {Li}\ \emph {et~al.}(2007)\citenamefont {Li}, \citenamefont {Wang}, \citenamefont {Sun},\ and\ \citenamefont {Yi}}]{li2007quantum}%
  \BibitemOpen
  \bibfield  {author} {\bibinfo {author} {\bibfnamefont {S.}~\bibnamefont {Li}}, \bibinfo {author} {\bibfnamefont {H.}~\bibnamefont {Wang}}, \bibinfo {author} {\bibfnamefont {Y.}~\bibnamefont {Sun}},\ and\ \bibinfo {author} {\bibfnamefont {X.}~\bibnamefont {Yi}},\ }\bibfield  {title} {\bibinfo {title} {Quantum heat engine beyond the adiabatic approximation},\ }\href {https://doi.org/10.1088/1751-8113/40/30/004} {\bibfield  {journal} {\bibinfo  {journal} {Journal of Physics A: Mathematical and Theoretical}\ }\textbf {\bibinfo {volume} {40}},\ \bibinfo {pages} {8655} (\bibinfo {year} {2007})}\BibitemShut {NoStop}%
\bibitem [{\citenamefont {Hardal}\ and\ \citenamefont {M{\"u}stecapl{\i}o{\u{g}}lu}(2015)}]{hardal2015superradiant}%
  \BibitemOpen
  \bibfield  {author} {\bibinfo {author} {\bibfnamefont {A.~{\"U}.}\ \bibnamefont {Hardal}}\ and\ \bibinfo {author} {\bibfnamefont {{\"O}.~E.}\ \bibnamefont {M{\"u}stecapl{\i}o{\u{g}}lu}},\ }\bibfield  {title} {\bibinfo {title} {Superradiant quantum heat engine},\ }\href {https://doi.org/10.1038/srep12953} {\bibfield  {journal} {\bibinfo  {journal} {Scientific reports}\ }\textbf {\bibinfo {volume} {5}},\ \bibinfo {pages} {12953} (\bibinfo {year} {2015})}\BibitemShut {NoStop}%
\bibitem [{\citenamefont {Oladimeji}\ \emph {et~al.}(2024)\citenamefont {Oladimeji}, \citenamefont {Idundun}, \citenamefont {Umeh}, \citenamefont {Ibrahim}, \citenamefont {Ikot}, \citenamefont {Koffa},\ and\ \citenamefont {Audu}}]{oladimeji2024performance}%
  \BibitemOpen
  \bibfield  {author} {\bibinfo {author} {\bibfnamefont {E.}~\bibnamefont {Oladimeji}}, \bibinfo {author} {\bibfnamefont {V.}~\bibnamefont {Idundun}}, \bibinfo {author} {\bibfnamefont {E.}~\bibnamefont {Umeh}}, \bibinfo {author} {\bibfnamefont {T.}~\bibnamefont {Ibrahim}}, \bibinfo {author} {\bibfnamefont {A.}~\bibnamefont {Ikot}}, \bibinfo {author} {\bibfnamefont {J.}~\bibnamefont {Koffa}},\ and\ \bibinfo {author} {\bibfnamefont {J.}~\bibnamefont {Audu}},\ }\bibfield  {title} {\bibinfo {title} {The performance analysis of a quantum mechanical carnot-like engine using diatomic molecules},\ }\href {https://doi.org/10.1007/s10909-024-03114-0} {\bibfield  {journal} {\bibinfo  {journal} {Journal of Low Temperature Physics}\ }\textbf {\bibinfo {volume} {216}},\ \bibinfo {pages} {538} (\bibinfo {year} {2024})}\BibitemShut {NoStop}%
\bibitem [{\citenamefont {Oladimeji}(2019)}]{oladimeji2019efficiency}%
  \BibitemOpen
  \bibfield  {author} {\bibinfo {author} {\bibfnamefont {E.}~\bibnamefont {Oladimeji}},\ }\bibfield  {title} {\bibinfo {title} {The efficiency of quantum engines using the p{\"o}schl--teller like oscillator model},\ }\href {https://doi.org/10.1016/j.physe.2019.03.002} {\bibfield  {journal} {\bibinfo  {journal} {Physica E: Low-Dimensional Systems and Nanostructures}\ }\textbf {\bibinfo {volume} {111}},\ \bibinfo {pages} {113} (\bibinfo {year} {2019})}\BibitemShut {NoStop}%
\bibitem [{\citenamefont {Oladimeji}\ \emph {et~al.}(2021)\citenamefont {Oladimeji}, \citenamefont {Owolabi},\ and\ \citenamefont {Adeleke}}]{oladimeji2021poschl}%
  \BibitemOpen
  \bibfield  {author} {\bibinfo {author} {\bibfnamefont {E.}~\bibnamefont {Oladimeji}}, \bibinfo {author} {\bibfnamefont {S.}~\bibnamefont {Owolabi}},\ and\ \bibinfo {author} {\bibfnamefont {J.}~\bibnamefont {Adeleke}},\ }\bibfield  {title} {\bibinfo {title} {The p{\"o}schl-teller like description of quantum-mechanical carnot engine},\ }\href {https://doi.org/10.1016/j.cjph.2021.01.004} {\bibfield  {journal} {\bibinfo  {journal} {Chinese Journal of Physics}\ }\textbf {\bibinfo {volume} {70}},\ \bibinfo {pages} {151} (\bibinfo {year} {2021})}\BibitemShut {NoStop}%
\bibitem [{\citenamefont {Ro{\ss}nagel}\ \emph {et~al.}(2016)\citenamefont {Ro{\ss}nagel}, \citenamefont {Dawkins}, \citenamefont {Tolazzi}, \citenamefont {Abah}, \citenamefont {Lutz}, \citenamefont {Schmidt-Kaler},\ and\ \citenamefont {Singer}}]{rossnagel2016single}%
  \BibitemOpen
  \bibfield  {author} {\bibinfo {author} {\bibfnamefont {J.}~\bibnamefont {Ro{\ss}nagel}}, \bibinfo {author} {\bibfnamefont {S.~T.}\ \bibnamefont {Dawkins}}, \bibinfo {author} {\bibfnamefont {K.~N.}\ \bibnamefont {Tolazzi}}, \bibinfo {author} {\bibfnamefont {O.}~\bibnamefont {Abah}}, \bibinfo {author} {\bibfnamefont {E.}~\bibnamefont {Lutz}}, \bibinfo {author} {\bibfnamefont {F.}~\bibnamefont {Schmidt-Kaler}},\ and\ \bibinfo {author} {\bibfnamefont {K.}~\bibnamefont {Singer}},\ }\bibfield  {title} {\bibinfo {title} {A single-atom heat engine},\ }\href {https://doi.org/10.1126/science.aad6320} {\bibfield  {journal} {\bibinfo  {journal} {Science}\ }\textbf {\bibinfo {volume} {352}},\ \bibinfo {pages} {325} (\bibinfo {year} {2016})}\BibitemShut {NoStop}%
\bibitem [{\citenamefont {Blickle}\ and\ \citenamefont {Bechinger}(2012)}]{blickle2012realization}%
  \BibitemOpen
  \bibfield  {author} {\bibinfo {author} {\bibfnamefont {V.}~\bibnamefont {Blickle}}\ and\ \bibinfo {author} {\bibfnamefont {C.}~\bibnamefont {Bechinger}},\ }\bibfield  {title} {\bibinfo {title} {Realization of a micrometre-sized stochastic heat engine},\ }\href {https://doi.org/https://doi.org/10.1038/nphys2163} {\bibfield  {journal} {\bibinfo  {journal} {Nature Physics}\ }\textbf {\bibinfo {volume} {8}},\ \bibinfo {pages} {143} (\bibinfo {year} {2012})}\BibitemShut {NoStop}%
\bibitem [{\citenamefont {Von~Lindenfels}\ \emph {et~al.}(2019)\citenamefont {Von~Lindenfels}, \citenamefont {Gr{\"a}b}, \citenamefont {Schmiegelow}, \citenamefont {Kaushal}, \citenamefont {Schulz}, \citenamefont {Mitchison}, \citenamefont {Goold}, \citenamefont {Schmidt-Kaler},\ and\ \citenamefont {Poschinger}}]{von2019spin}%
  \BibitemOpen
  \bibfield  {author} {\bibinfo {author} {\bibfnamefont {D.}~\bibnamefont {Von~Lindenfels}}, \bibinfo {author} {\bibfnamefont {O.}~\bibnamefont {Gr{\"a}b}}, \bibinfo {author} {\bibfnamefont {C.~T.}\ \bibnamefont {Schmiegelow}}, \bibinfo {author} {\bibfnamefont {V.}~\bibnamefont {Kaushal}}, \bibinfo {author} {\bibfnamefont {J.}~\bibnamefont {Schulz}}, \bibinfo {author} {\bibfnamefont {M.~T.}\ \bibnamefont {Mitchison}}, \bibinfo {author} {\bibfnamefont {J.}~\bibnamefont {Goold}}, \bibinfo {author} {\bibfnamefont {F.}~\bibnamefont {Schmidt-Kaler}},\ and\ \bibinfo {author} {\bibfnamefont {U.~G.}\ \bibnamefont {Poschinger}},\ }\bibfield  {title} {\bibinfo {title} {Spin heat engine coupled to a harmonic-oscillator flywheel},\ }\href {https://doi.org/https://doi.org/10.1103/PhysRevLett.123.080602} {\bibfield  {journal} {\bibinfo  {journal} {Physical Review Letters}\ }\textbf {\bibinfo {volume} {123}},\ \bibinfo {pages} {080602} (\bibinfo {year} {2019})}\BibitemShut {NoStop}%
\bibitem [{\citenamefont {Peterson}\ \emph {et~al.}(2019)\citenamefont {Peterson}, \citenamefont {Batalhao}, \citenamefont {Herrera}, \citenamefont {Souza}, \citenamefont {Sarthour}, \citenamefont {Oliveira},\ and\ \citenamefont {Serra}}]{peterson2019experimental}%
  \BibitemOpen
  \bibfield  {author} {\bibinfo {author} {\bibfnamefont {J.~P.}\ \bibnamefont {Peterson}}, \bibinfo {author} {\bibfnamefont {T.~B.}\ \bibnamefont {Batalhao}}, \bibinfo {author} {\bibfnamefont {M.}~\bibnamefont {Herrera}}, \bibinfo {author} {\bibfnamefont {A.~M.}\ \bibnamefont {Souza}}, \bibinfo {author} {\bibfnamefont {R.~S.}\ \bibnamefont {Sarthour}}, \bibinfo {author} {\bibfnamefont {I.~S.}\ \bibnamefont {Oliveira}},\ and\ \bibinfo {author} {\bibfnamefont {R.~M.}\ \bibnamefont {Serra}},\ }\bibfield  {title} {\bibinfo {title} {Experimental characterization of a spin quantum heat engine},\ }\href {https://doi.org/https://doi.org/10.1103/PhysRevLett.123.240601} {\bibfield  {journal} {\bibinfo  {journal} {Physical Review Letters}\ }\textbf {\bibinfo {volume} {123}},\ \bibinfo {pages} {240601} (\bibinfo {year} {2019})}\BibitemShut {NoStop}%
\bibitem [{\citenamefont {Klatzow}\ \emph {et~al.}(2019)\citenamefont {Klatzow}, \citenamefont {Becker}, \citenamefont {Ledingham}, \citenamefont {Weinzetl}, \citenamefont {Kaczmarek}, \citenamefont {Saunders}, \citenamefont {Nunn}, \citenamefont {Walmsley}, \citenamefont {Uzdin},\ and\ \citenamefont {Poem}}]{klatzow2019experimental}%
  \BibitemOpen
  \bibfield  {author} {\bibinfo {author} {\bibfnamefont {J.}~\bibnamefont {Klatzow}}, \bibinfo {author} {\bibfnamefont {J.~N.}\ \bibnamefont {Becker}}, \bibinfo {author} {\bibfnamefont {P.~M.}\ \bibnamefont {Ledingham}}, \bibinfo {author} {\bibfnamefont {C.}~\bibnamefont {Weinzetl}}, \bibinfo {author} {\bibfnamefont {K.~T.}\ \bibnamefont {Kaczmarek}}, \bibinfo {author} {\bibfnamefont {D.~J.}\ \bibnamefont {Saunders}}, \bibinfo {author} {\bibfnamefont {J.}~\bibnamefont {Nunn}}, \bibinfo {author} {\bibfnamefont {I.~A.}\ \bibnamefont {Walmsley}}, \bibinfo {author} {\bibfnamefont {R.}~\bibnamefont {Uzdin}},\ and\ \bibinfo {author} {\bibfnamefont {E.}~\bibnamefont {Poem}},\ }\bibfield  {title} {\bibinfo {title} {Experimental demonstration of quantum effects in the operation of microscopic heat engines},\ }\href {https://doi.org/https://doi.org/10.1103/PhysRevLett.122.110601} {\bibfield  {journal} {\bibinfo  {journal} {Physical Review Letters}\ }\textbf {\bibinfo {volume} {122}},\ \bibinfo {pages} {110601}
  (\bibinfo {year} {2019})}\BibitemShut {NoStop}%
\bibitem [{\citenamefont {Passos}\ \emph {et~al.}(2019)\citenamefont {Passos}, \citenamefont {Santos}, \citenamefont {Sarandy},\ and\ \citenamefont {Huguenin}}]{passos2019optical}%
  \BibitemOpen
  \bibfield  {author} {\bibinfo {author} {\bibfnamefont {M.}~\bibnamefont {Passos}}, \bibinfo {author} {\bibfnamefont {A.~C.}\ \bibnamefont {Santos}}, \bibinfo {author} {\bibfnamefont {M.~S.}\ \bibnamefont {Sarandy}},\ and\ \bibinfo {author} {\bibfnamefont {J.}~\bibnamefont {Huguenin}},\ }\bibfield  {title} {\bibinfo {title} {Optical simulation of a quantum thermal machine},\ }\href {https://doi.org/https://doi.org/10.1103/PhysRevA.100.022113} {\bibfield  {journal} {\bibinfo  {journal} {Physical Review A}\ }\textbf {\bibinfo {volume} {100}},\ \bibinfo {pages} {022113} (\bibinfo {year} {2019})}\BibitemShut {NoStop}%
\bibitem [{\citenamefont {Morse}(1929)}]{morse1929diatomic}%
  \BibitemOpen
  \bibfield  {author} {\bibinfo {author} {\bibfnamefont {P.~M.}\ \bibnamefont {Morse}},\ }\bibfield  {title} {\bibinfo {title} {Diatomic molecules according to the wave mechanics. ii. vibrational levels},\ }\href {https://doi.org/10.1103/PhysRev.34.57} {\bibfield  {journal} {\bibinfo  {journal} {Physical Review}\ }\textbf {\bibinfo {volume} {34}},\ \bibinfo {pages} {57} (\bibinfo {year} {1929})}\BibitemShut {NoStop}%
\bibitem [{\citenamefont {Schmidt}\ and\ \citenamefont {Steel}(2024)}]{schmidt2024molecular}%
  \BibitemOpen
  \bibfield  {author} {\bibinfo {author} {\bibfnamefont {M.~K.}\ \bibnamefont {Schmidt}}\ and\ \bibinfo {author} {\bibfnamefont {M.~J.}\ \bibnamefont {Steel}},\ }\bibfield  {title} {\bibinfo {title} {Molecular optomechanics in the anharmonic regime: from nonclassical mechanical states to mechanical lasing},\ }\href {https://doi.org/10.1088/1367-2630/ad32e4} {\bibfield  {journal} {\bibinfo  {journal} {New Journal of Physics}\ }\textbf {\bibinfo {volume} {26}},\ \bibinfo {pages} {033041} (\bibinfo {year} {2024})}\BibitemShut {NoStop}%
\bibitem [{\citenamefont {Krantz}\ \emph {et~al.}(2019)\citenamefont {Krantz}, \citenamefont {Kjaergaard}, \citenamefont {Yan}, \citenamefont {Orlando}, \citenamefont {Gustavsson},\ and\ \citenamefont {Oliver}}]{krantz2019quantum}%
  \BibitemOpen
  \bibfield  {author} {\bibinfo {author} {\bibfnamefont {P.}~\bibnamefont {Krantz}}, \bibinfo {author} {\bibfnamefont {M.}~\bibnamefont {Kjaergaard}}, \bibinfo {author} {\bibfnamefont {F.}~\bibnamefont {Yan}}, \bibinfo {author} {\bibfnamefont {T.~P.}\ \bibnamefont {Orlando}}, \bibinfo {author} {\bibfnamefont {S.}~\bibnamefont {Gustavsson}},\ and\ \bibinfo {author} {\bibfnamefont {W.~D.}\ \bibnamefont {Oliver}},\ }\bibfield  {title} {\bibinfo {title} {A quantum engineer's guide to superconducting qubits},\ }\href {https://doi.org/10.1063/1.5089550} {\bibfield  {journal} {\bibinfo  {journal} {Applied Physics Reviews}\ }\textbf {\bibinfo {volume} {6}} (\bibinfo {year} {2019})}\BibitemShut {NoStop}%
\bibitem [{\citenamefont {Home}\ \emph {et~al.}(2011)\citenamefont {Home}, \citenamefont {Hanneke}, \citenamefont {Jost}, \citenamefont {Leibfried},\ and\ \citenamefont {Wineland}}]{home2011normal}%
  \BibitemOpen
  \bibfield  {author} {\bibinfo {author} {\bibfnamefont {J.~P.}\ \bibnamefont {Home}}, \bibinfo {author} {\bibfnamefont {D.}~\bibnamefont {Hanneke}}, \bibinfo {author} {\bibfnamefont {J.~D.}\ \bibnamefont {Jost}}, \bibinfo {author} {\bibfnamefont {D.}~\bibnamefont {Leibfried}},\ and\ \bibinfo {author} {\bibfnamefont {D.~J.}\ \bibnamefont {Wineland}},\ }\bibfield  {title} {\bibinfo {title} {Normal modes of trapped ions in the presence of anharmonic trap potentials},\ }\href {https://doi.org/10.1088/1367-2630/13/7/073026} {\bibfield  {journal} {\bibinfo  {journal} {New Journal of Physics}\ }\textbf {\bibinfo {volume} {13}},\ \bibinfo {pages} {073026} (\bibinfo {year} {2011})}\BibitemShut {NoStop}%
\bibitem [{\citenamefont {Galperin}\ \emph {et~al.}(2007)\citenamefont {Galperin}, \citenamefont {Ratner},\ and\ \citenamefont {Nitzan}}]{galperin2007molecular}%
  \BibitemOpen
  \bibfield  {author} {\bibinfo {author} {\bibfnamefont {M.}~\bibnamefont {Galperin}}, \bibinfo {author} {\bibfnamefont {M.~A.}\ \bibnamefont {Ratner}},\ and\ \bibinfo {author} {\bibfnamefont {A.}~\bibnamefont {Nitzan}},\ }\bibfield  {title} {\bibinfo {title} {Molecular transport junctions: vibrational effects},\ }\href {https://doi.org/10.1088/0953-8984/19/10/103201} {\bibfield  {journal} {\bibinfo  {journal} {Journal of Physics: Condensed Matter}\ }\textbf {\bibinfo {volume} {19}},\ \bibinfo {pages} {103201} (\bibinfo {year} {2007})}\BibitemShut {NoStop}%
\bibitem [{\citenamefont {L{\"u}}\ \emph {et~al.}(2015)\citenamefont {L{\"u}}, \citenamefont {Zhou}, \citenamefont {Jiang},\ and\ \citenamefont {Wang}}]{lu2015effects}%
  \BibitemOpen
  \bibfield  {author} {\bibinfo {author} {\bibfnamefont {J.-T.}\ \bibnamefont {L{\"u}}}, \bibinfo {author} {\bibfnamefont {H.}~\bibnamefont {Zhou}}, \bibinfo {author} {\bibfnamefont {J.-W.}\ \bibnamefont {Jiang}},\ and\ \bibinfo {author} {\bibfnamefont {J.-S.}\ \bibnamefont {Wang}},\ }\bibfield  {title} {\bibinfo {title} {Effects of electron-phonon interaction on thermal and electrical transport through molecular nano-conductors},\ }\href {https://doi.org/10.1063/1.4917017} {\bibfield  {journal} {\bibinfo  {journal} {AIP Advances}\ }\textbf {\bibinfo {volume} {5}} (\bibinfo {year} {2015})}\BibitemShut {NoStop}%
\bibitem [{\citenamefont {L{\"u}}\ \emph {et~al.}(2011)\citenamefont {L{\"u}}, \citenamefont {Hedeg{\aa}rd},\ and\ \citenamefont {Brandbyge}}]{lu2011laserlike}%
  \BibitemOpen
  \bibfield  {author} {\bibinfo {author} {\bibfnamefont {J.-T.}\ \bibnamefont {L{\"u}}}, \bibinfo {author} {\bibfnamefont {P.}~\bibnamefont {Hedeg{\aa}rd}},\ and\ \bibinfo {author} {\bibfnamefont {M.}~\bibnamefont {Brandbyge}},\ }\bibfield  {title} {\bibinfo {title} {Laserlike vibrational instability in rectifying molecular conductors},\ }\href {https://doi.org/10.1103/PhysRevLett.107.046801} {\bibfield  {journal} {\bibinfo  {journal} {Physical Review Letters}\ }\textbf {\bibinfo {volume} {107}},\ \bibinfo {pages} {046801} (\bibinfo {year} {2011})}\BibitemShut {NoStop}%
\bibitem [{\citenamefont {Simine}\ and\ \citenamefont {Segal}(2012)}]{simine2012vibrational}%
  \BibitemOpen
  \bibfield  {author} {\bibinfo {author} {\bibfnamefont {L.}~\bibnamefont {Simine}}\ and\ \bibinfo {author} {\bibfnamefont {D.}~\bibnamefont {Segal}},\ }\bibfield  {title} {\bibinfo {title} {Vibrational cooling, heating, and instability in molecular conducting junctions: full counting statistics analysis},\ }\href {https://doi.org/10.1039/C2CP40851A} {\bibfield  {journal} {\bibinfo  {journal} {Physical Chemistry Chemical Physics}\ }\textbf {\bibinfo {volume} {14}},\ \bibinfo {pages} {13820} (\bibinfo {year} {2012})}\BibitemShut {NoStop}%
\bibitem [{\citenamefont {Simine}\ and\ \citenamefont {Segal}(2013)}]{simine2013path}%
  \BibitemOpen
  \bibfield  {author} {\bibinfo {author} {\bibfnamefont {L.}~\bibnamefont {Simine}}\ and\ \bibinfo {author} {\bibfnamefont {D.}~\bibnamefont {Segal}},\ }\bibfield  {title} {\bibinfo {title} {Path-integral simulations with fermionic and bosonic reservoirs: Transport and dissipation in molecular electronic junctions},\ }\href {https://doi.org/10.1063/1.4808108} {\bibfield  {journal} {\bibinfo  {journal} {The Journal of Chemical Physics}\ }\textbf {\bibinfo {volume} {138}} (\bibinfo {year} {2013})}\BibitemShut {NoStop}%
\bibitem [{\citenamefont {Arrachea}\ \emph {et~al.}(2014)\citenamefont {Arrachea}, \citenamefont {Bode},\ and\ \citenamefont {Von~Oppen}}]{arrachea2014vibrational}%
  \BibitemOpen
  \bibfield  {author} {\bibinfo {author} {\bibfnamefont {L.}~\bibnamefont {Arrachea}}, \bibinfo {author} {\bibfnamefont {N.}~\bibnamefont {Bode}},\ and\ \bibinfo {author} {\bibfnamefont {F.}~\bibnamefont {Von~Oppen}},\ }\bibfield  {title} {\bibinfo {title} {Vibrational cooling and thermoelectric response of nanoelectromechanical systems},\ }\href {https://doi.org/10.1103/PhysRevB.90.125450} {\bibfield  {journal} {\bibinfo  {journal} {Physical Review B}\ }\textbf {\bibinfo {volume} {90}},\ \bibinfo {pages} {125450} (\bibinfo {year} {2014})}\BibitemShut {NoStop}%
\bibitem [{\citenamefont {Agarwalla}\ and\ \citenamefont {Segal}(2016)}]{agarwalla2016reconciling}%
  \BibitemOpen
  \bibfield  {author} {\bibinfo {author} {\bibfnamefont {B.~K.}\ \bibnamefont {Agarwalla}}\ and\ \bibinfo {author} {\bibfnamefont {D.}~\bibnamefont {Segal}},\ }\bibfield  {title} {\bibinfo {title} {Reconciling perturbative approaches in phonon-assisted transport junctions},\ }\href {https://doi.org/10.1063/1.4941582} {\bibfield  {journal} {\bibinfo  {journal} {The Journal of Chemical Physics}\ }\textbf {\bibinfo {volume} {144}} (\bibinfo {year} {2016})}\BibitemShut {NoStop}%
\bibitem [{\citenamefont {Erpenbeck}\ \emph {et~al.}(2015)\citenamefont {Erpenbeck}, \citenamefont {Haertle},\ and\ \citenamefont {Thoss}}]{erpenbeck2015effect}%
  \BibitemOpen
  \bibfield  {author} {\bibinfo {author} {\bibfnamefont {A.}~\bibnamefont {Erpenbeck}}, \bibinfo {author} {\bibfnamefont {R.}~\bibnamefont {Haertle}},\ and\ \bibinfo {author} {\bibfnamefont {M.}~\bibnamefont {Thoss}},\ }\bibfield  {title} {\bibinfo {title} {Effect of nonadiabatic electronic-vibrational interactions on the transport properties of single-molecule junctions},\ }\href {https://doi.org/10.1103/PhysRevB.91.195418} {\bibfield  {journal} {\bibinfo  {journal} {Physical Review B}\ }\textbf {\bibinfo {volume} {91}},\ \bibinfo {pages} {195418} (\bibinfo {year} {2015})}\BibitemShut {NoStop}%
\bibitem [{\citenamefont {Friedman}\ \emph {et~al.}(2017)\citenamefont {Friedman}, \citenamefont {Agarwalla},\ and\ \citenamefont {Segal}}]{friedman2017effects}%
  \BibitemOpen
  \bibfield  {author} {\bibinfo {author} {\bibfnamefont {H.~M.}\ \bibnamefont {Friedman}}, \bibinfo {author} {\bibfnamefont {B.~K.}\ \bibnamefont {Agarwalla}},\ and\ \bibinfo {author} {\bibfnamefont {D.}~\bibnamefont {Segal}},\ }\bibfield  {title} {\bibinfo {title} {Effects of vibrational anharmonicity on molecular electronic conduction and thermoelectric efficiency},\ }\href {https://doi.org/10.1063/1.4965824} {\bibfield  {journal} {\bibinfo  {journal} {The Journal of Chemical Physics}\ }\textbf {\bibinfo {volume} {146}} (\bibinfo {year} {2017})}\BibitemShut {NoStop}%
\bibitem [{\citenamefont {Xomalis}\ \emph {et~al.}(2021)\citenamefont {Xomalis}, \citenamefont {Zheng}, \citenamefont {Chikkaraddy}, \citenamefont {Koczor-Benda}, \citenamefont {Miele}, \citenamefont {Rosta}, \citenamefont {Vandenbosch}, \citenamefont {Mart{\'\i}nez},\ and\ \citenamefont {Baumberg}}]{xomalis2021detecting}%
  \BibitemOpen
  \bibfield  {author} {\bibinfo {author} {\bibfnamefont {A.}~\bibnamefont {Xomalis}}, \bibinfo {author} {\bibfnamefont {X.}~\bibnamefont {Zheng}}, \bibinfo {author} {\bibfnamefont {R.}~\bibnamefont {Chikkaraddy}}, \bibinfo {author} {\bibfnamefont {Z.}~\bibnamefont {Koczor-Benda}}, \bibinfo {author} {\bibfnamefont {E.}~\bibnamefont {Miele}}, \bibinfo {author} {\bibfnamefont {E.}~\bibnamefont {Rosta}}, \bibinfo {author} {\bibfnamefont {G.~A.}\ \bibnamefont {Vandenbosch}}, \bibinfo {author} {\bibfnamefont {A.}~\bibnamefont {Mart{\'\i}nez}},\ and\ \bibinfo {author} {\bibfnamefont {J.~J.}\ \bibnamefont {Baumberg}},\ }\bibfield  {title} {\bibinfo {title} {Detecting mid-infrared light by molecular frequency upconversion in dual-wavelength nanoantennas},\ }\href {https://doi.org/10.1126/science.abk25} {\bibfield  {journal} {\bibinfo  {journal} {Science}\ }\textbf {\bibinfo {volume} {374}},\ \bibinfo {pages} {1268} (\bibinfo {year} {2021})}\BibitemShut {NoStop}%
\bibitem [{\citenamefont {Chen}\ \emph {et~al.}(2021)\citenamefont {Chen}, \citenamefont {Roelli}, \citenamefont {Hu}, \citenamefont {Verlekar}, \citenamefont {Amirtharaj}, \citenamefont {Barreda}, \citenamefont {Kippenberg}, \citenamefont {Kovylina}, \citenamefont {Verhagen}, \citenamefont {Mart{\'\i}nez} \emph {et~al.}}]{chen2021continuous}%
  \BibitemOpen
  \bibfield  {author} {\bibinfo {author} {\bibfnamefont {W.}~\bibnamefont {Chen}}, \bibinfo {author} {\bibfnamefont {P.}~\bibnamefont {Roelli}}, \bibinfo {author} {\bibfnamefont {H.}~\bibnamefont {Hu}}, \bibinfo {author} {\bibfnamefont {S.}~\bibnamefont {Verlekar}}, \bibinfo {author} {\bibfnamefont {S.~P.}\ \bibnamefont {Amirtharaj}}, \bibinfo {author} {\bibfnamefont {A.~I.}\ \bibnamefont {Barreda}}, \bibinfo {author} {\bibfnamefont {T.~J.}\ \bibnamefont {Kippenberg}}, \bibinfo {author} {\bibfnamefont {M.}~\bibnamefont {Kovylina}}, \bibinfo {author} {\bibfnamefont {E.}~\bibnamefont {Verhagen}}, \bibinfo {author} {\bibfnamefont {A.}~\bibnamefont {Mart{\'\i}nez}}, \emph {et~al.},\ }\bibfield  {title} {\bibinfo {title} {Continuous-wave frequency upconversion with a molecular optomechanical nanocavity},\ }\href {https://doi.org/10.1126/science.abk3106} {\bibfield  {journal} {\bibinfo  {journal} {Science}\ }\textbf {\bibinfo {volume} {374}},\ \bibinfo {pages} {1264} (\bibinfo {year} {2021})}\BibitemShut {NoStop}%
\bibitem [{\citenamefont {Lombardi}\ \emph {et~al.}(2018)\citenamefont {Lombardi}, \citenamefont {Schmidt}, \citenamefont {Weller}, \citenamefont {Deacon}, \citenamefont {Benz}, \citenamefont {De~Nijs}, \citenamefont {Aizpurua},\ and\ \citenamefont {Baumberg}}]{lombardi2018pulsed}%
  \BibitemOpen
  \bibfield  {author} {\bibinfo {author} {\bibfnamefont {A.}~\bibnamefont {Lombardi}}, \bibinfo {author} {\bibfnamefont {M.~K.}\ \bibnamefont {Schmidt}}, \bibinfo {author} {\bibfnamefont {L.}~\bibnamefont {Weller}}, \bibinfo {author} {\bibfnamefont {W.~M.}\ \bibnamefont {Deacon}}, \bibinfo {author} {\bibfnamefont {F.}~\bibnamefont {Benz}}, \bibinfo {author} {\bibfnamefont {B.}~\bibnamefont {De~Nijs}}, \bibinfo {author} {\bibfnamefont {J.}~\bibnamefont {Aizpurua}},\ and\ \bibinfo {author} {\bibfnamefont {J.~J.}\ \bibnamefont {Baumberg}},\ }\bibfield  {title} {\bibinfo {title} {Pulsed molecular optomechanics in plasmonic nanocavities: from nonlinear vibrational instabilities to bond-breaking},\ }\href {https://doi.org/10.1103/PhysRevX.8.011016} {\bibfield  {journal} {\bibinfo  {journal} {Physical Review X}\ }\textbf {\bibinfo {volume} {8}},\ \bibinfo {pages} {011016} (\bibinfo {year} {2018})}\BibitemShut {NoStop}%
\bibitem [{\citenamefont {Kosloff}\ and\ \citenamefont {Rezek}(2017)}]{Kosloff2017}%
  \BibitemOpen
  \bibfield  {author} {\bibinfo {author} {\bibfnamefont {R.}~\bibnamefont {Kosloff}}\ and\ \bibinfo {author} {\bibfnamefont {Y.}~\bibnamefont {Rezek}},\ }\bibfield  {title} {\bibinfo {title} {The quantum harmonic otto cycle},\ }\href {https://www.mdpi.com/1099-4300/19/4/136} {\bibfield  {journal} {\bibinfo  {journal} {Entropy}\ }\textbf {\bibinfo {volume} {19}} (\bibinfo {year} {2017})}\BibitemShut {NoStop}%
\bibitem [{\citenamefont {Arai}(1991)}]{arai1991exactly}%
  \BibitemOpen
  \bibfield  {author} {\bibinfo {author} {\bibfnamefont {A.}~\bibnamefont {Arai}},\ }\bibfield  {title} {\bibinfo {title} {Exactly solvable supersymmetric quantum mechanics},\ }\href {https://doi.org/10.1016/0022-247X(91)90267-4} {\bibfield  {journal} {\bibinfo  {journal} {Journal of Mathematical Analysis and Applications}\ }\textbf {\bibinfo {volume} {158}},\ \bibinfo {pages} {63} (\bibinfo {year} {1991})}\BibitemShut {NoStop}%
\bibitem [{\citenamefont {Arai}(2001)}]{arai2001exact}%
  \BibitemOpen
  \bibfield  {author} {\bibinfo {author} {\bibfnamefont {A.}~\bibnamefont {Arai}},\ }\bibfield  {title} {\bibinfo {title} {Exact solutions ofmulti-component nonlinear schr{\"o}dinger and klein-gordonequations in two-dimensional space-time},\ }\href {https://doi.org/10.1088/0305-4470/34/20/302} {\bibfield  {journal} {\bibinfo  {journal} {Journal of Physics A: Mathematical and General}\ }\textbf {\bibinfo {volume} {34}},\ \bibinfo {pages} {4281} (\bibinfo {year} {2001})}\BibitemShut {NoStop}%
\bibitem [{\citenamefont {R-Monteiro}\ \emph {et~al.}(1996)\citenamefont {R-Monteiro}, \citenamefont {Rodrigues},\ and\ \citenamefont {Wulck}}]{r1996quantum}%
  \BibitemOpen
  \bibfield  {author} {\bibinfo {author} {\bibfnamefont {M.}~\bibnamefont {R-Monteiro}}, \bibinfo {author} {\bibfnamefont {L.}~\bibnamefont {Rodrigues}},\ and\ \bibinfo {author} {\bibfnamefont {S.}~\bibnamefont {Wulck}},\ }\bibfield  {title} {\bibinfo {title} {Quantum algebraic nature of the phonon spectrum in 4 he},\ }\href {https://doi.org/10.1103/PhysRevLett.76.1098} {\bibfield  {journal} {\bibinfo  {journal} {Physical Review Letters}\ }\textbf {\bibinfo {volume} {76}},\ \bibinfo {pages} {1098} (\bibinfo {year} {1996})}\BibitemShut {NoStop}%
\bibitem [{\citenamefont {Bonatsos}\ \emph {et~al.}(1997)\citenamefont {Bonatsos}, \citenamefont {Daskaloyannis},\ and\ \citenamefont {Kolokotronis}}]{bonatsos1997coupled}%
  \BibitemOpen
  \bibfield  {author} {\bibinfo {author} {\bibfnamefont {D.}~\bibnamefont {Bonatsos}}, \bibinfo {author} {\bibfnamefont {C.}~\bibnamefont {Daskaloyannis}},\ and\ \bibinfo {author} {\bibfnamefont {P.}~\bibnamefont {Kolokotronis}},\ }\bibfield  {title} {\bibinfo {title} {Coupled q-oscillators as a model for vibrations of polyatomic molecules},\ }\href {https://doi.org/10.1063/1.473189} {\bibfield  {journal} {\bibinfo  {journal} {The Journal of Chemical Physics}\ }\textbf {\bibinfo {volume} {106}},\ \bibinfo {pages} {605} (\bibinfo {year} {1997})}\BibitemShut {NoStop}%
\bibitem [{\citenamefont {Johal}\ and\ \citenamefont {Gupta}(1998)}]{johal1998two}%
  \BibitemOpen
  \bibfield  {author} {\bibinfo {author} {\bibfnamefont {R.~S.}\ \bibnamefont {Johal}}\ and\ \bibinfo {author} {\bibfnamefont {R.~K.}\ \bibnamefont {Gupta}},\ }\bibfield  {title} {\bibinfo {title} {Two parameter quantum deformation of u (2) $\supset$ u (1) dynamical symmetry and the vibrational spectra of diatomic molecules},\ }\href {https://doi.org/10.1142/S0218301398000294} {\bibfield  {journal} {\bibinfo  {journal} {International Journal of Modern Physics E}\ }\textbf {\bibinfo {volume} {7}},\ \bibinfo {pages} {553} (\bibinfo {year} {1998})}\BibitemShut {NoStop}%
\bibitem [{\citenamefont {Alavi}\ and\ \citenamefont {Rouhani}(2004)}]{alavi2004exact}%
  \BibitemOpen
  \bibfield  {author} {\bibinfo {author} {\bibfnamefont {S.}~\bibnamefont {Alavi}}\ and\ \bibinfo {author} {\bibfnamefont {S.}~\bibnamefont {Rouhani}},\ }\bibfield  {title} {\bibinfo {title} {Exact analytical expression for magnetoresistance using quantum groups},\ }\href {https://doi.org/10.1016/j.physleta.2003.11.028} {\bibfield  {journal} {\bibinfo  {journal} {Physics Letters A}\ }\textbf {\bibinfo {volume} {320}},\ \bibinfo {pages} {327} (\bibinfo {year} {2004})}\BibitemShut {NoStop}%
\bibitem [{\citenamefont {Cooper}\ and\ \citenamefont {Gupta}(1995)}]{cooper1995q}%
  \BibitemOpen
  \bibfield  {author} {\bibinfo {author} {\bibfnamefont {I.~L.}\ \bibnamefont {Cooper}}\ and\ \bibinfo {author} {\bibfnamefont {R.~K.}\ \bibnamefont {Gupta}},\ }\bibfield  {title} {\bibinfo {title} {q-deformed morse oscillator},\ }\href {https://doi.org/10.1103/PhysRevA.52.941} {\bibfield  {journal} {\bibinfo  {journal} {Physical Review A}\ }\textbf {\bibinfo {volume} {52}},\ \bibinfo {pages} {941} (\bibinfo {year} {1995})}\BibitemShut {NoStop}%
\bibitem [{\citenamefont {Ikhdair}(2009)}]{ikhdair2009rotation}%
  \BibitemOpen
  \bibfield  {author} {\bibinfo {author} {\bibfnamefont {S.~M.}\ \bibnamefont {Ikhdair}},\ }\bibfield  {title} {\bibinfo {title} {Rotation and vibration of diatomic molecule in the spatially-dependent mass schr{\"o}dinger equation with generalized q-deformed morse potential},\ }\href {https://doi.org/https://doi.org/10.1016/j.chemphys.2009.04.023} {\bibfield  {journal} {\bibinfo  {journal} {Chemical Physics}\ }\textbf {\bibinfo {volume} {361}},\ \bibinfo {pages} {9} (\bibinfo {year} {2009})}\BibitemShut {NoStop}%
\bibitem [{\citenamefont {Dobrogowska}(2013)}]{dobrogowska2013q}%
  \BibitemOpen
  \bibfield  {author} {\bibinfo {author} {\bibfnamefont {A.}~\bibnamefont {Dobrogowska}},\ }\bibfield  {title} {\bibinfo {title} {The q-deformation of the morse potential},\ }\href {https://doi.org/10.1016/j.aml.2013.02.009} {\bibfield  {journal} {\bibinfo  {journal} {Applied Mathematics Letters}\ }\textbf {\bibinfo {volume} {26}},\ \bibinfo {pages} {769} (\bibinfo {year} {2013})}\BibitemShut {NoStop}%
\bibitem [{\citenamefont {Nutku}\ \emph {et~al.}(2022)\citenamefont {Nutku}, \citenamefont {Aydiner},\ and\ \citenamefont {Sen}}]{nutku2022complexity}%
  \BibitemOpen
  \bibfield  {author} {\bibinfo {author} {\bibfnamefont {F.}~\bibnamefont {Nutku}}, \bibinfo {author} {\bibfnamefont {E.}~\bibnamefont {Aydiner}},\ and\ \bibinfo {author} {\bibfnamefont {K.}~\bibnamefont {Sen}},\ }\bibfield  {title} {\bibinfo {title} {Complexity of hcl and h2 molecules under q-deformed morse potential},\ }\href {https://doi.org/https://doi.org/10.1007/s12648-021-02028-x} {\bibfield  {journal} {\bibinfo  {journal} {Indian Journal of Physics}\ }\textbf {\bibinfo {volume} {96}},\ \bibinfo {pages} {1} (\bibinfo {year} {2022})}\BibitemShut {NoStop}%
\bibitem [{\citenamefont {Akta{\c{s}}}\ and\ \citenamefont {Sever}(2004)}]{aktacs2004supersymmetric}%
  \BibitemOpen
  \bibfield  {author} {\bibinfo {author} {\bibfnamefont {M.}~\bibnamefont {Akta{\c{s}}}}\ and\ \bibinfo {author} {\bibfnamefont {R.}~\bibnamefont {Sever}},\ }\bibfield  {title} {\bibinfo {title} {Supersymmetric solution of pt-/non-pt-symmetric and non-hermitian morse potential via hamiltonian hierarchy method},\ }\href {https://doi.org/10.1142/S0217732304015993} {\bibfield  {journal} {\bibinfo  {journal} {Modern Physics Letters A}\ }\textbf {\bibinfo {volume} {19}},\ \bibinfo {pages} {2871} (\bibinfo {year} {2004})}\BibitemShut {NoStop}%
\bibitem [{\citenamefont {Hassanabadi}\ \emph {et~al.}(2018)\citenamefont {Hassanabadi}, \citenamefont {Sargolzaeipor},\ and\ \citenamefont {Chung}}]{hassanabadi2018superstatistics}%
  \BibitemOpen
  \bibfield  {author} {\bibinfo {author} {\bibfnamefont {H.}~\bibnamefont {Hassanabadi}}, \bibinfo {author} {\bibfnamefont {S.}~\bibnamefont {Sargolzaeipor}},\ and\ \bibinfo {author} {\bibfnamefont {W.}~\bibnamefont {Chung}},\ }\bibfield  {title} {\bibinfo {title} {Superstatistics properties of q-deformed morse potential in one dimension},\ }\href {https://doi.org/10.1016/j.physa.2018.05.125} {\bibfield  {journal} {\bibinfo  {journal} {Physica A: Statistical Mechanics and its Applications}\ }\textbf {\bibinfo {volume} {508}},\ \bibinfo {pages} {740} (\bibinfo {year} {2018})}\BibitemShut {NoStop}%
\bibitem [{\citenamefont {Boumali}(2018)}]{boumali2018statistical}%
  \BibitemOpen
  \bibfield  {author} {\bibinfo {author} {\bibfnamefont {A.}~\bibnamefont {Boumali}},\ }\bibfield  {title} {\bibinfo {title} {The statistical properties of q-deformed morse potential for some diatomic molecules via euler--maclaurin method in one dimension},\ }\href {https://doi.org/10.1007/s10910-018-0879-4} {\bibfield  {journal} {\bibinfo  {journal} {Journal of Mathematical Chemistry}\ }\textbf {\bibinfo {volume} {56}},\ \bibinfo {pages} {1656} (\bibinfo {year} {2018})}\BibitemShut {NoStop}%
\bibitem [{\citenamefont {Leonard}\ and\ \citenamefont {Deffner}(2015)}]{leonard2015quantum}%
  \BibitemOpen
  \bibfield  {author} {\bibinfo {author} {\bibfnamefont {A.}~\bibnamefont {Leonard}}\ and\ \bibinfo {author} {\bibfnamefont {S.}~\bibnamefont {Deffner}},\ }\bibfield  {title} {\bibinfo {title} {Quantum work distribution for a driven diatomic molecule},\ }\href {https://doi.org/10.1016/j.chemphys.2014.10.020} {\bibfield  {journal} {\bibinfo  {journal} {Chemical Physics}\ }\textbf {\bibinfo {volume} {446}},\ \bibinfo {pages} {18} (\bibinfo {year} {2015})}\BibitemShut {NoStop}%
\bibitem [{\citenamefont {Beck}\ and\ \citenamefont {Cohen}(2003)}]{beck2003superstatistics}%
  \BibitemOpen
  \bibfield  {author} {\bibinfo {author} {\bibfnamefont {C.}~\bibnamefont {Beck}}\ and\ \bibinfo {author} {\bibfnamefont {E.~G.}\ \bibnamefont {Cohen}},\ }\bibfield  {title} {\bibinfo {title} {Superstatistics},\ }\href {https://doi.org/10.1016/S0378-4371(03)00019-0} {\bibfield  {journal} {\bibinfo  {journal} {Physica A: Statistical mechanics and its applications}\ }\textbf {\bibinfo {volume} {322}},\ \bibinfo {pages} {267} (\bibinfo {year} {2003})}\BibitemShut {NoStop}%
\bibitem [{\citenamefont {Quan}\ \emph {et~al.}(2005{\natexlab{b}})\citenamefont {Quan}, \citenamefont {Zhang},\ and\ \citenamefont {Sun}}]{quan2005quantum}%
  \BibitemOpen
  \bibfield  {author} {\bibinfo {author} {\bibfnamefont {H.}~\bibnamefont {Quan}}, \bibinfo {author} {\bibfnamefont {P.}~\bibnamefont {Zhang}},\ and\ \bibinfo {author} {\bibfnamefont {C.}~\bibnamefont {Sun}},\ }\bibfield  {title} {\bibinfo {title} {Quantum heat engine with multilevel quantum systems},\ }\href {https://doi.org/https://doi.org/10.1103/PhysRevE.72.056110} {\bibfield  {journal} {\bibinfo  {journal} {Physical Review E}\ }\textbf {\bibinfo {volume} {72}},\ \bibinfo {pages} {056110} (\bibinfo {year} {2005}{\natexlab{b}})}\BibitemShut {NoStop}%
\bibitem [{\citenamefont {Quan}(2009)}]{quan2009quantum}%
  \BibitemOpen
  \bibfield  {author} {\bibinfo {author} {\bibfnamefont {H.~T.}\ \bibnamefont {Quan}},\ }\bibfield  {title} {\bibinfo {title} {Quantum thermodynamic cycles and quantum heat engines. ii.},\ }\href {https://doi.org/https://doi.org/10.1103/PhysRevE.79.041129} {\bibfield  {journal} {\bibinfo  {journal} {Physical Review E}\ }\textbf {\bibinfo {volume} {79}},\ \bibinfo {pages} {041129} (\bibinfo {year} {2009})}\BibitemShut {NoStop}%
\bibitem [{\citenamefont {Prakash}\ \emph {et~al.}(2022)\citenamefont {Prakash}, \citenamefont {Kumar},\ and\ \citenamefont {Benjamin}}]{Prakash2022}%
  \BibitemOpen
  \bibfield  {author} {\bibinfo {author} {\bibfnamefont {A.}~\bibnamefont {Prakash}}, \bibinfo {author} {\bibfnamefont {A.}~\bibnamefont {Kumar}},\ and\ \bibinfo {author} {\bibfnamefont {C.}~\bibnamefont {Benjamin}},\ }\bibfield  {title} {\bibinfo {title} {Impurity reveals distinct operational phases in quantum thermodynamic cycles},\ }\href {https://link.aps.org/doi/10.1103/PhysRevE.106.054112} {\bibfield  {journal} {\bibinfo  {journal} {Phys. Rev. E}\ }\textbf {\bibinfo {volume} {106}},\ \bibinfo {pages} {054112} (\bibinfo {year} {2022})}\BibitemShut {NoStop}%
\bibitem [{\citenamefont {Pal}\ and\ \citenamefont {Benjamin}(2022)}]{pal2022josephson}%
  \BibitemOpen
  \bibfield  {author} {\bibinfo {author} {\bibfnamefont {S.}~\bibnamefont {Pal}}\ and\ \bibinfo {author} {\bibfnamefont {C.}~\bibnamefont {Benjamin}},\ }\bibfield  {title} {\bibinfo {title} {Josephson quantum spin thermodynamics},\ }\href {https://doi.org/10.1088/1361-648X/ac6f3b} {\bibfield  {journal} {\bibinfo  {journal} {Journal of Physics: Condensed Matter}\ }\textbf {\bibinfo {volume} {34}},\ \bibinfo {pages} {305601} (\bibinfo {year} {2022})}\BibitemShut {NoStop}%
\bibitem [{\citenamefont {de~Oliveira}\ \emph {et~al.}(2021)\citenamefont {de~Oliveira}, \citenamefont {Rojas},\ and\ \citenamefont {Filgueiras}}]{de2021two}%
  \BibitemOpen
  \bibfield  {author} {\bibinfo {author} {\bibfnamefont {J.~L.~D.}\ \bibnamefont {de~Oliveira}}, \bibinfo {author} {\bibfnamefont {M.}~\bibnamefont {Rojas}},\ and\ \bibinfo {author} {\bibfnamefont {C.}~\bibnamefont {Filgueiras}},\ }\bibfield  {title} {\bibinfo {title} {Two coupled double quantum-dot systems as a working substance for heat machines},\ }\href {https://doi.org/10.1103/PhysRevE.104.014149} {\bibfield  {journal} {\bibinfo  {journal} {Physical Review E}\ }\textbf {\bibinfo {volume} {104}},\ \bibinfo {pages} {014149} (\bibinfo {year} {2021})}\BibitemShut {NoStop}%
\bibitem [{\citenamefont {Ozaydin}\ \emph {et~al.}(2023)\citenamefont {Ozaydin}, \citenamefont {M{\"u}stecapl{\i}o{\u{g}}lu},\ and\ \citenamefont {Hakio{\u{g}}lu}}]{ozaydin2023powering}%
  \BibitemOpen
  \bibfield  {author} {\bibinfo {author} {\bibfnamefont {F.}~\bibnamefont {Ozaydin}}, \bibinfo {author} {\bibfnamefont {{\"O}.~E.}\ \bibnamefont {M{\"u}stecapl{\i}o{\u{g}}lu}},\ and\ \bibinfo {author} {\bibfnamefont {T.}~\bibnamefont {Hakio{\u{g}}lu}},\ }\bibfield  {title} {\bibinfo {title} {Powering quantum otto engines only with q-deformation of the working substance},\ }\href {https://doi.org/10.1103/PhysRevE.108.054103} {\bibfield  {journal} {\bibinfo  {journal} {Physical Review E}\ }\textbf {\bibinfo {volume} {108}},\ \bibinfo {pages} {054103} (\bibinfo {year} {2023})}\BibitemShut {NoStop}%
\bibitem [{\citenamefont {Gelbwaser-Klimovsky}\ \emph {et~al.}(2018)\citenamefont {Gelbwaser-Klimovsky}, \citenamefont {Bylinskii}, \citenamefont {Gangloff}, \citenamefont {Islam}, \citenamefont {Aspuru-Guzik},\ and\ \citenamefont {Vuletic}}]{gelbwaser2018single}%
  \BibitemOpen
  \bibfield  {author} {\bibinfo {author} {\bibfnamefont {D.}~\bibnamefont {Gelbwaser-Klimovsky}}, \bibinfo {author} {\bibfnamefont {A.}~\bibnamefont {Bylinskii}}, \bibinfo {author} {\bibfnamefont {D.}~\bibnamefont {Gangloff}}, \bibinfo {author} {\bibfnamefont {R.}~\bibnamefont {Islam}}, \bibinfo {author} {\bibfnamefont {A.}~\bibnamefont {Aspuru-Guzik}},\ and\ \bibinfo {author} {\bibfnamefont {V.}~\bibnamefont {Vuletic}},\ }\bibfield  {title} {\bibinfo {title} {Single-atom heat machines enabled by energy quantization},\ }\href {https://doi.org/10.1103/PhysRevLett.120.170601} {\bibfield  {journal} {\bibinfo  {journal} {Physical Review Letters}\ }\textbf {\bibinfo {volume} {120}},\ \bibinfo {pages} {170601} (\bibinfo {year} {2018})}\BibitemShut {NoStop}%
\end{thebibliography}%
\end{document}